\documentclass[a4paper,11pt]{article}
\pdfoutput=1 

\usepackage{jcappub} 

\usepackage[T1]{fontenc} 
\usepackage[dvipsnames]{xcolor}

\title{\boldmath Spectrum of Cuscuton Bounce}

\author[a,b,c]{J. Leo Kim}
\author[a,b,c]{and Ghazal Geshnizjani}

\affiliation[a]{Department of Applied Mathematics, University of Waterloo \\200 University Ave W,  N2L 3G1, Waterloo, Ontario, Canada}
\affiliation[b]{Perimeter Institute for Theoretical Physics \\31 Caroline St. N., N2L 2Y5, Waterloo, Ontario,  Canada}
\affiliation[c]{Waterloo Centre for Astrophysics, University of Waterloo \\200 University Ave W,  N2L 3G1, Waterloo, Ontario, Canada}

\emailAdd{jlkim@uwaterloo.ca}
\emailAdd{ggeshniz@uwaterloo.ca}

\abstract{It has been recently shown that a cosmological bounce model based on Cuscuton gravity does not have any ghosts or curvature instabilities. We explore whether Cuscuton bounce can provide an alternative to inflation for generating near scale-invariant scalar perturbations. While a single field Cuscuton bounce generically produces a strongly blue power spectrum (for a variety of initial/boundary conditions), we demonstrate that scale-invariant entropy modes can be generated in a spectator field that starts in adiabatic vacuum, and is kinetically coupled to the primary field. Furthermore, our solution has no singularity, nor requires an {\it ad hoc} matching condition. We also study the generation of tensor modes (or gravitational waves) in Cuscuton bounce and show that while they are stable, similar to other bounce models, the produced spectrum is strongly blue and unobservable.}

\begin{document}
\maketitle
\flushbottom

\section{Introduction}\label{intro}

Over the past forty years, the inflationary paradigm \cite{Guth:1980zm, Linde:1981mu} has gradually become the widely accepted theory to describe the initial conditions of the Universe. Originally motivated to mainly address the flatness and horizon problems, inflationary models also provide an impressive mechanism to generate seeds of fluctuations in the gravitational background on cosmological scales (length scales of gigaparsecs, $\sim 10^{60} \ell_P$) from vacuum quantum fluctuations on extremely small scales (length scales $\lesssim H^{-1} \approx 10^{-19} \text{\normalfont\AA} \sim 10^{6} \ell_P$) \cite{Senatore:2016aui}. As the scales corresponding to these fluctuations cross the Hubble horizon during inflation, their power spectra become nearly scale-invariant ($n_s\simeq 1$, $n_t\simeq 0$) for both scalar modes and tensor modes, and they also have coherent phases which leads to acoustic peaks in the temperature, polarization, and matter power spectra. This near scale-invariance prediction is consistent with cosmological observations such as the Cosmic Microwave Background (CMB) measurements by WMAP \cite{Hinshaw:2003ex} and Planck\footnote{We note that the Planck collaboration observed a spectral tilt of $n_s = 0.965\pm 0.004$, which is $8\sigma$ away from scale invariance ($n_s = 1$). However, from the model building point of view the more significant development is getting $n_s$ close to one. In the language of inflation that translates into first realizing a de Sitter back-ground geometry that produces exact scale invariant perturbations, then by slightly deviating from it and allowing a varying potential rather than a cosmological constant produce the deviation from one.} \cite{Aghanim:2018eyx} as well as the Baryon Acoustic Oscillation (BAO) in galaxy survey power spectra (e.g., \cite{Alam:2020sor}). In addition, the precision measurement of cosmological parameters such as the value of spectral index $n_s$ can be used to exclude or constrain many inflationary models \cite{Akrami:2018odb}. 

From the theoretical perspective, there is no consensus on a theory beyond the Standard Model of particle physics that can be used to verify or eliminate inflation as a viable theory. Still, one can test the validity of the effective field theory and semi-classical gravity assumptions that are used in inflation and their consistency with current theoretical candidates to describe quantum gravity. For example while the effective field theory itself and loop corrections during inflation seem to be under control \cite{Babic:2019ify,Burgess:2017ytm}, it is still important to understand whether or not inflation faces issues such as the trans-Planckian problem \cite{Martin:2000xs, Brandenberger:2000wr, Brandenberger:2012aj}. The latter can arise from the period of inflation being possibly too long such that the physical wavelengths of cosmological observations today correspond to sub-Planck length scales at the beginning of inflation. If that is the case then the Bunch–Davies vacuum may not be the correct initial conditions and the theory needs to be further extended in order to adequately set the initial conditions \cite{Ashoorioon:2018uey}. Therefore, it is still not clear whether inflation provides a fully self-consistent theoretical framework for setting initial conditions in the early Universe. 

More recently, the trans-Planckian problem has been promoted to the Trans-Planckian Censorship Conjecture (TCC) \cite{Bedroya:2019snp, Bedroya:2019tba}, positing that models that encounter the trans-Planckian problem are inconsistent with string theory, i.e. they lie in the ``swampland'' \cite{Ooguri:2006in, Brennan:2017rbf, Obied:2018sgi, Palti:2019pca}. The swampland conjectures provides a list of criteria for effective field theories that can arise from string theory. However, constraints on inflationary models from observational data are believed to be in strong tension with the swampland conjectures \cite{Achucarro:2018vey, Dias:2018ngv, Agrawal:2018own, Kinney:2018nny, Lin:2019pmj}. Regardless of whether the swampland conjectures are to be believed, the question of fine-tuning/naturalness has also been a matter of much contention in the cosmology communities. In the language of smoothing \cite{Garfinkle:2008ei, Cook:2020oaj, Ijjas:2020dws}, the dynamical attractor solution for inflation is a flat, homogeneous and isotropic universe, but it has been argued that is not necessarily a quantum smoother \cite{steinhardt1982natural, Vilenkin:1983xq}. In addition to these, inflationary spacetimes are also eternal in nature \cite{Linde:1986fc, Guth:2007ng} and seem to be geodesically past incomplete, which implies they do not address the singularity problem \cite{Borde:1993xh}. To summarize, taking everything into account from theory to observations, one could argue that in spite substantial circumstantial evidence for an early phase of inflation \cite{Geshnizjani:2011dk, Geshnizjani:2013lza, Geshnizjani:2014bya}, its compatibility within a larger theoretical high energy physics framework is far from certain. With this being said, the trans-Planckian problem is an ongoing topic of discussion with arguments for both side of the debate (for a relatively recent argument against the trans-Planckian problem, see \cite{Dvali:2020cgt}).

With all this in mind, it is natural to ask whether alternative models of the Early Universe can avoid these suggested shortcomings and/or be less contrived. One natural alternative that could directly address the horizon and singularity problems is a bounce scenario in which the Universe initially undergoes a contracting phase, pauses momentarily and then proceeds to enter an expansion phase. Over the years, many different bouncing models have also been proposed and studied in detail \cite{Finelli:2001sr, Brandenberger:1988aj, Qiu:2011cy, Cai:2007zv, Cai:2012va, Cai:2013kja, Gasperini:1992em, Cai:2016thi, Cai:2017tku, Cai:2017dyi, Lehners:2008vx, Khoury:2001wf, Ijjas:2016vtq, Ijjas:2016tpn,Dobre:2017pnt, Easson:2011zy, Creminelli:2006xe, Creminelli:2007aq, Creminelli:2016zwa}, each with their own set of defining characteristics and obstacles. These bouncing models can be classified into two main categories, either singular bounce models or regular (non-singular) bounce scenarios. In particular, the regular bouncing cosmologies have a finite curvature and energy density at the bounce, naturally addressing the singularity problem. However, these models generically violate the Null Energy Condition (NEC) $\rho + p \geq 0$, which in general relativity and most theories of modified gravity leads to either instabilities or a superluminal speed of sound \cite{Dubovsky:2005xd, Sawicki:2012pz, Rubakov:2014jja, Libanov:2016kfc,  Kobayashi:2016xpl}.\footnote{A superluminal propagation speed may not necessary imply that causality is violated \cite{Babichev:2007dw}. However, for configurations that allow superluminality, UV completeness can also be an issue \cite{Adams:2006sv}. For further discussion on superluminality, we refer readers to some of the many papers discussing this topic and the references therein \cite{Easson:2013bda, Libanov:2016kfc, Babichev:2007dw, Adams:2006sv, Dobre:2017pnt, Mironov:2019haz, Mironov:2019mye, Mironov:2020pqh}.} Remarkably, it was shown recently that a bouncing cosmology generated by Cuscuton gravity \cite{Boruah:2017tvg, Boruah:2018pvq} can work around all these difficulties. Cuscuton gravity \cite{Afshordi:2006ad, Afshordi:2007yx} is an infrared modification to gravity, which is implemented though an auxiliary field without its own dynamical degrees of freedom. Similar to general relativity, in order to induce a dynamical cosmological background, one has to include other matter fields. At its simplest form, this auxiliary field is a non-canonical scalar field that is incompressible. The studies in \cite{Boruah:2017tvg, Boruah:2018pvq} show that a Cuscuton bounce does not have any ghost instabilities and the scalar perturbations remain stable throughout the bounce phase. This crucial result relies on the fact that, while providing a mechanism for the bounce, the Cuscuton field does not have any dynamical degrees of freedom.\footnote{This has been shown for Cuscuton without coupling to gravity and pertubatively around cosmological backgrounds. However, \cite{Gomes:2017tzd} suggests that a generic inhomogeneous initial condition for Cuscuton may lead to a propagating degree of freedom. However, whether under such conditions the equations remain well-posed and system is still physical or not is not clear.} This allows for an effective violation of the NEC for the background while the matter fields remain safe. The stability results for these perturbations were explored in further detail \cite{Quintin:2019orx} for both Cuscuton gravity as well its extended version \cite{Iyonaga:2018vnu}. 
Note that a bounce scenario need not exclude an inflationary phase. For example, a bouncing universe followed by an inflationary phase can address the singularity problem and also generate the seeds for inhomogeneities in the universe. However, even a more interesting possibility is to see if there is a way to address both of these aspects without requiring an inflationary phase.   

The goal of this paper is mainly to investigate that second possibility, i.e. study the power spectrum of scalar perturbations in a Cuscuton bounce cosmology. We start with section \eqref{review} which provides a review of the Cuscuton bounce scenario presented in \cite{Boruah:2018pvq} and the reason this model is free of scalar instabilities throughout the bounce. Next, we show in section \eqref{tensor} why the tensor perturbations are also free of instabilities through the bounce. The power spectra for scalar perturbations in single field Cuscuton bounce model are explored in section \eqref{PSsingle}, where we argue despite various initial conditions, the adiabatic cosmological perturbations cannot produce nearly scale-invariant power spectra. We will then show in section \eqref{twofield} that by adding another scalar field that is kinetically coupled to the primary matter field, a near scale-invariant power spectrum can be obtained for entropy perturbations, either before or after the bounce phase. Finally, we return to the tensor perturbations in section \eqref{PStensor} and show that they produce an unobservably small, but strongly blue power spectrum assuming adiabatic vacuum initial condition. We end this paper by making our concluding remarks in section \eqref{conclusion}.


\section{The single field Cuscuton bounce scenario} \label{review}

In this section, we will review the Cuscuton bounce model and the stability studies that were carried out in \cite{Boruah:2018pvq}. The general action, including the Einstein-Hilbert term, the kinetic and potential terms for Cuscuton field $\varphi$, and a dynamical scalar field $\pi$, with minimal coupling and no potential is given by:
\begin{align} 
    S = \int d^4 x \sqrt{-g} \left[ \frac{M_P^2}{2} R - \mu^2 \sqrt{-D_\mu \varphi D^\mu \varphi} - V(\varphi)- \frac{1}{2} D_\mu \pi D^\mu \pi  \right], \label{eq:oldaction}
\end{align}
where $D_\mu$ denotes the covariant derivative.\footnote{In general, the kinetic term for Cuscuton action can be taken to be negative or positive. However, we are using the negative sign since only that can induce a regular bounce solution.} As demonstrated in \cite{Boruah:2018pvq}, choosing a potential $V(\varphi)$ with some generic features around the bounce and far from it can induce a cosmological bounce solution. In this scenario, a potential consistent with those features was taken to be
\begin{figure}[tbp]
\centering 
\includegraphics[width=0.9\textwidth]{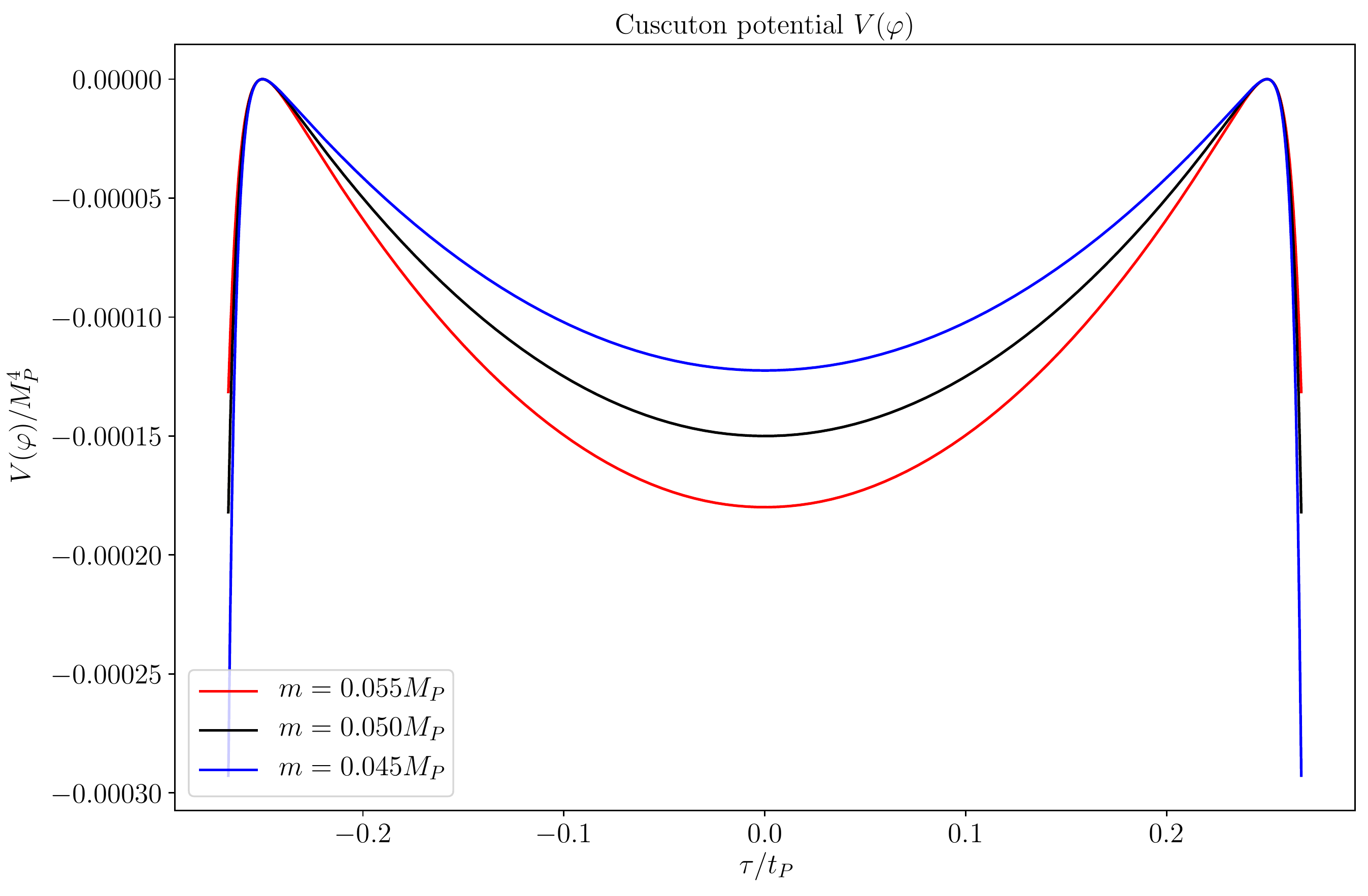}
\caption{\label{fig:V} Cuscuton potential $V(\varphi)$ for a range of $\varphi$. For this plot, $\varphi_\infty = 5m$ is fixed while $m$ is allowed to vary. The construction of this potential along with the selection of the necessary parameters are covered in detail in \cite{Boruah:2018pvq}.}
\end{figure}
\begin{align}
    V(\varphi) = m^2 (\varphi^2 - \varphi_\infty^2) - m^4 \left[ e^{(\varphi^2-\varphi_\infty^2) / m^2 } - 1 \right], \label{eq:V}
\end{align}
where $m,\mu$ and $\varphi_\infty$ are free parameters. Figure \eqref{fig:V} gives a visual intuition about the shape of the Cuscuton potential for different values of $\mu$ and a fixed $\varphi_\infty$. For the rest of the paper, we will fix values of $m = 0.05 M_P$ and $\varphi_\infty = 5m$, taking the same values as the original bounce paper. 

Considering a Friedmann–Lema\^itre–Robertson–Walker (FLRW) universe, variation of the action \eqref{eq:oldaction} with respect to the Cuscuton field $\varphi$ results in the following constraint equation:
\begin{align}\label{cusEOM}
    V'(\varphi_0)  - 3~ \text{sgn}(\dot{\varphi_0}) \mu^2 H  &= 0,
\end{align}
where $\varphi_0(t)$ is the homogeneous component of the $\varphi$-field. 
Without loss of generality, in what follows, we only consider self-consistent solutions where $\dot{\varphi_0}>0$\footnote{Note that our analysis is in the regime where $X=-D_\mu\varphi D^\mu \varphi>0$ is globally satisfied. Therefore, $\dot{\varphi_0}$ cannot be zero. For more discussion about the two different branches of the theory and the cosmological backgrounds, we refer the readers to the original Cuscuton papers \cite{Afshordi:2006ad, Afshordi:2007yx}.} This equation determines the relationship of the Cuscuton field $\varphi_0$ to the Hubble parameter $H$, and establishes that the sign of $V'(\varphi_0)$ will determine whether background is in contracting or expanding phase. Furthermore, as long as $V'(\varphi)$ is bounded $H$ will not diverge. Note that since there are no time derivatives of the Cuscuton field $\varphi$ in this equation, this confirms the lack of dynamical degrees of freedom at zeroth order in perturbations for $\varphi$ around a flat FRW background. Therefore, a homogeneous FRW background cannot evolve with just the Cuscuton field and another matter field with dynamical degrees of freedom (in this case taken to be $\pi$) is required. Since $\pi$ is a free field, its corresponding equation of motion for the background is given by
\begin{align}
    0 = 3 H \dot{\pi}_0 + \Ddot{\pi}_0, \label{eq:pieom}
\end{align}
where $\pi_0$ is the homogeneous component of the $\pi$-field. Note that it is from eq. \eqref{eq:pieom} that one can see that $\rho_\pi \propto a^{-6}$. This was done intentionally by setting the potential for the $\pi$-field to be zero, which results in the equation of state parameter $w \sim 1$. By doing this, the anisotropies will never take over the background density, which is one of the generic problems in other bounce models that lead to the BKL instability \cite{Belinsky:1970ew, Lifshitz:1963ps, Battefeld:2014uga}. Finally, the Einstein equations for the background lead to the following Friedmann equations:
\begin{align}
H^2 &= \frac{1}{3 M_P^2} \left[ \frac{1}{2} \dot{\pi}_0^2 + V(\varphi_0) \right] \label{eq:f1} \\
\dot{H} &= -\frac{1}{2M_P^2} \left[ - \mu^2 |\dot{\varphi}_0| + \dot{\pi}_0^2   \right] \label{eq:f2},
\end{align}
which when combined with eq. \eqref{cusEOM}, provides an expression for the time evolution for $\varphi$,
\begin{align}
    \dot{\varphi}_0 = \frac{ \frac{V(\varphi_0)}{M_P^2} - 3 H^2(\varphi_0) }{ \frac{V''(\varphi_0)}{3\mu^2}  - \frac{\mu^2}{2M_P^2} }.
\end{align}
In fact, by solving this first order ODE, all of the other dynamics in the background can be re-expressed in terms of $\varphi_0$, which then acts as a clock. This leads to very efficient numerical computations.

\begin{figure}[tbp]
\centering
\includegraphics[width=.47\textwidth]{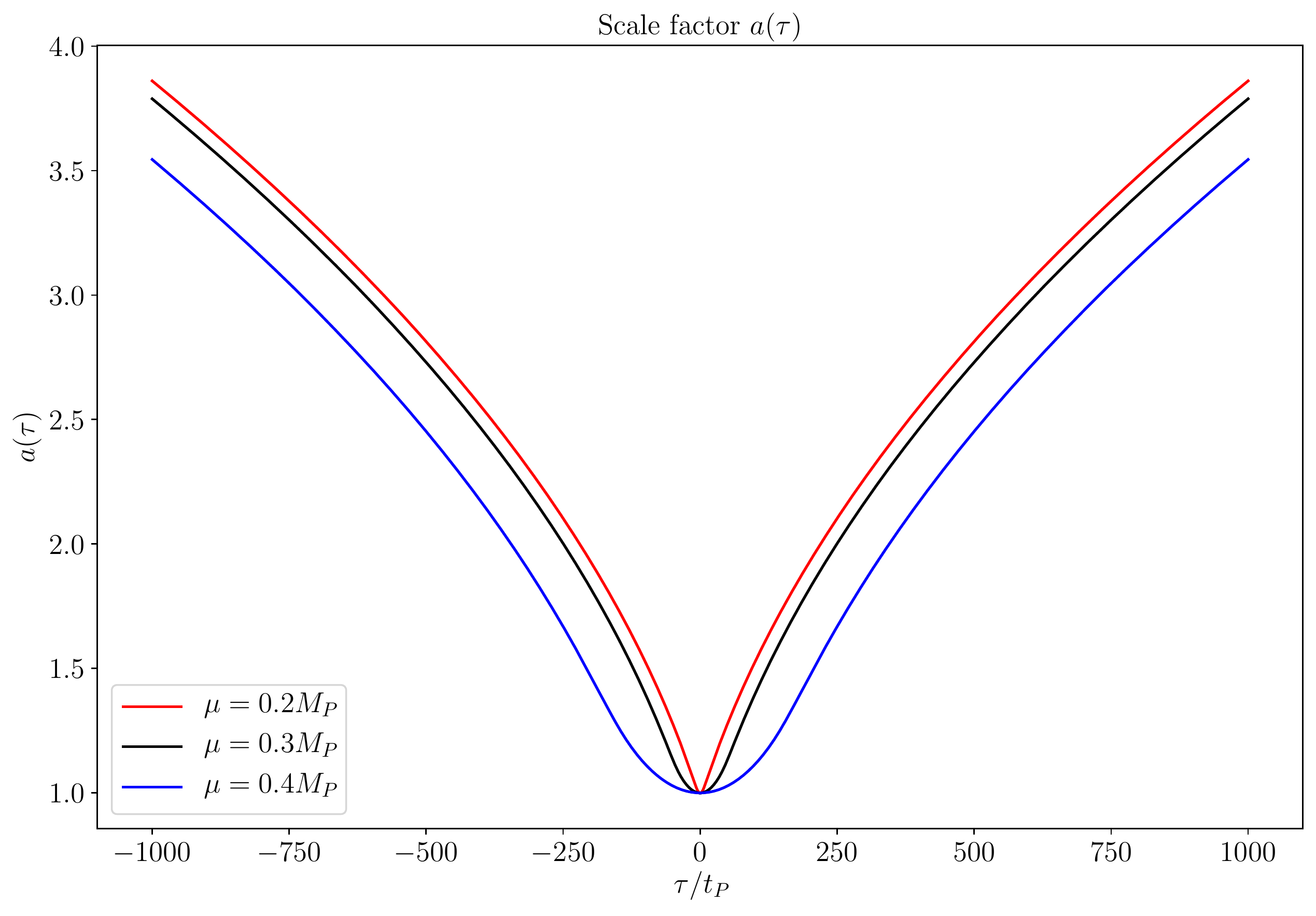}
\hfill
\includegraphics[width=.49\textwidth]{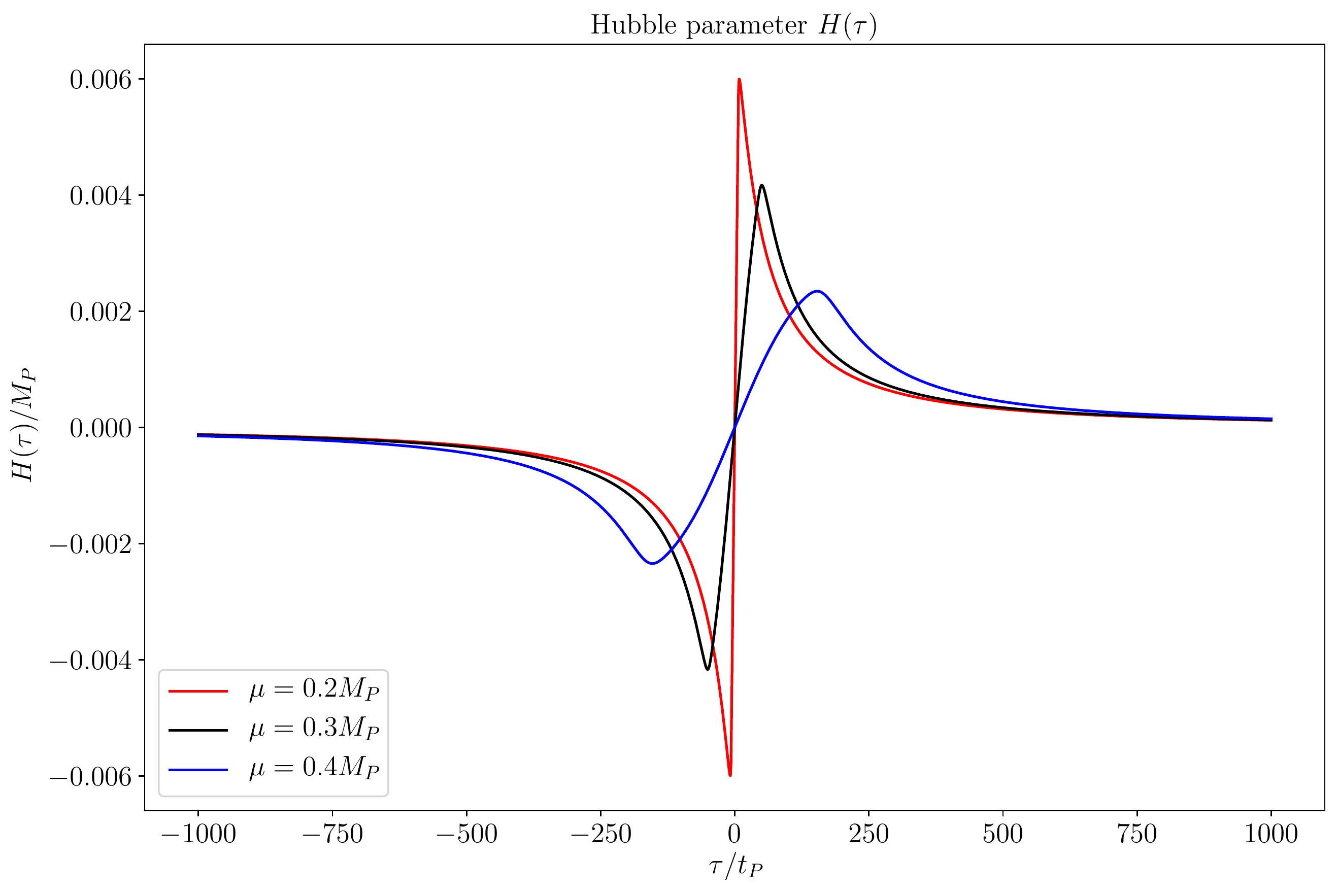}
\caption{\label{fig:background} Background quantities for Cuscuton bounce. The left figure is the scale factor $a(\tau)$ where the bounce was set at $\tau_b = 0$ and $a(\tau_b) = 1$ and $\tau$ denotes the conformal time. The figure on the right is the Hubble parameter $H(\tau)$.}
\end{figure}

Figure \eqref{fig:background} illustrates how the scale factor $a$ and Hubble parameter $H$ change with respect to time for different values of $\mu$ for this model. For what follows, we will also fix $\mu = 0.3 M_P$ to match with the original Cuscuton bounce paper \cite{Boruah:2018pvq}. 

With the background quantities established, the next step is to study cosmological perturbations in this model. Similar to the standard theory of cosmological perturbations, ADM formalism \cite{Arnowitt:1962hi, Poisson:2009pwt} can be applied here by splitting the $3+1$ spacetime into a spacelike foliation and a time direction, where the metric is given by:
\begin{align}
    ds^2 = - N^2 dt^2 + h_{ij} (dx^i + N^i dt) (dx^j + N^j dt).
\end{align}
Here $N$ and $N^i$ represent the lapse and shift functions while $h_{ij}$ is the induced metric on the three-dimensional spacelike hypersurfaces. After fixing one of the available gauge choices, $h_{ij}$ can be expressed in terms of the curvature perturbations $\zeta$ and the tensor perturbations $\gamma_{ij}$ as 
 \begin{align}
    h_{ij} = a^2[ \delta_{ij} (1 + 2 \zeta) + \gamma_{ij} ]. \label{eq:pert}
\end{align}
Expressing the perturbations for the Cuscuton field as $\varphi = \varphi_0 + \delta \varphi$ and the canonical scalar field as $\pi = \pi_0 + \delta \pi$, one can fix the leftover gauge choice by working in the unitary gauge for $\pi$ where $\delta \pi = 0$. Next, varying the action with respect to $N$ and $N^i$ leads to Hamiltonian and momentum constraints while varying with respect to $\delta \varphi$ provides the constraint equation for Cuscuton. Writing this equation in Fourier space, the relationship between $\zeta_k, \dot{\zeta}_k$, and $ \delta \varphi_k$ reduces to
\begin{align}
    \delta \varphi_k = \dot{\varphi_0} \frac{(k/a)^2 H \zeta_k + P \dot{\zeta}_k}{[ (k/a)^2 H^2 + P(3H^2 + P + \dot{H}) ]},
\end{align}
where $P = \frac{1}{2M_P^2} \dot{\pi}_0^2 $.\footnote{Note that since for simplicity in our calculations, we are setting $a=1$ at the bounce, the comoving wave number $k$ represents physical scale at the bounce and not the physical scale at present time.} Finally, thorough computation of action up to second order in perturbations in Fourier space yields
\begin{figure}[tbp]
\centering
\includegraphics[width=.49\textwidth]{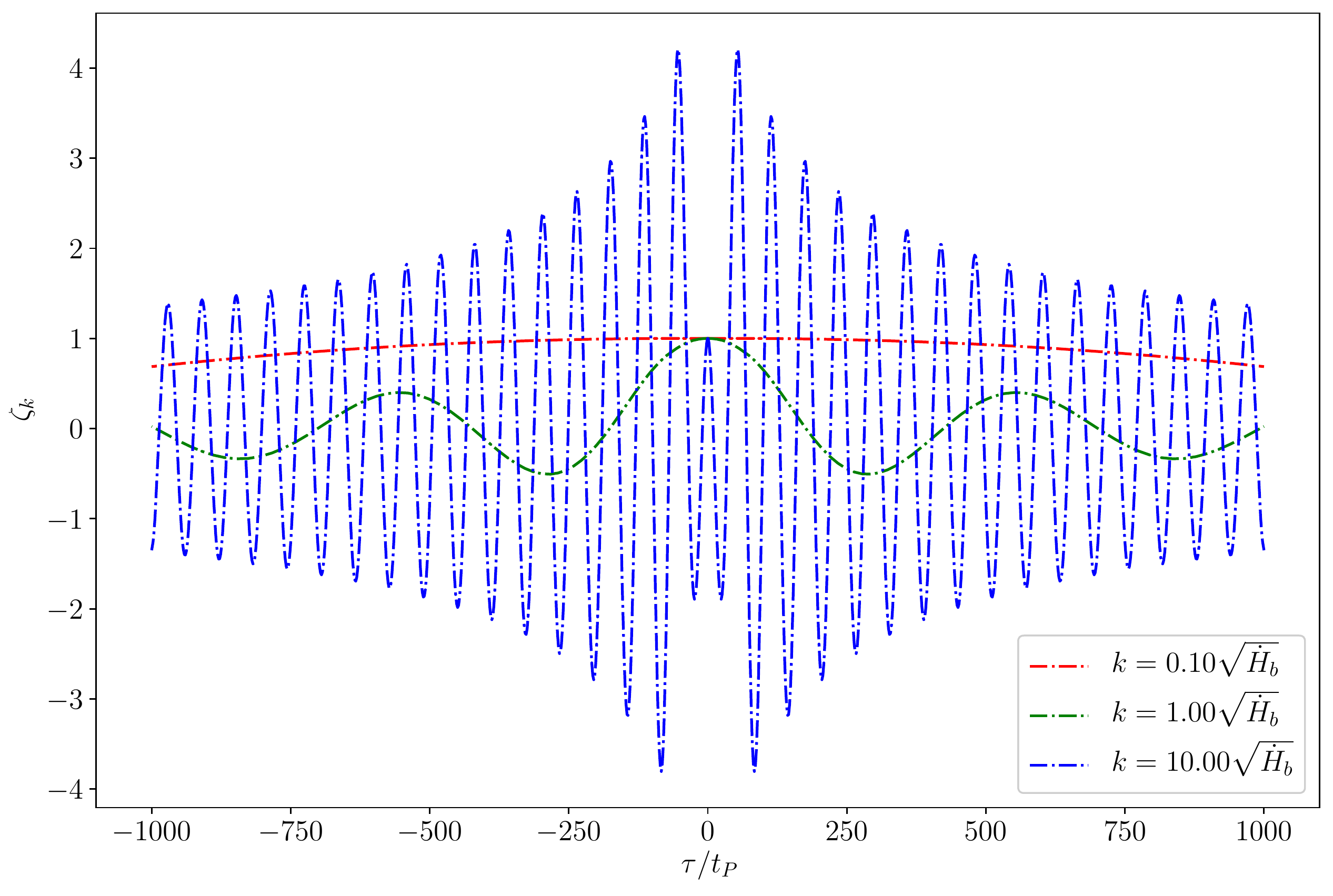}
\hfill
\includegraphics[width=.49\textwidth]{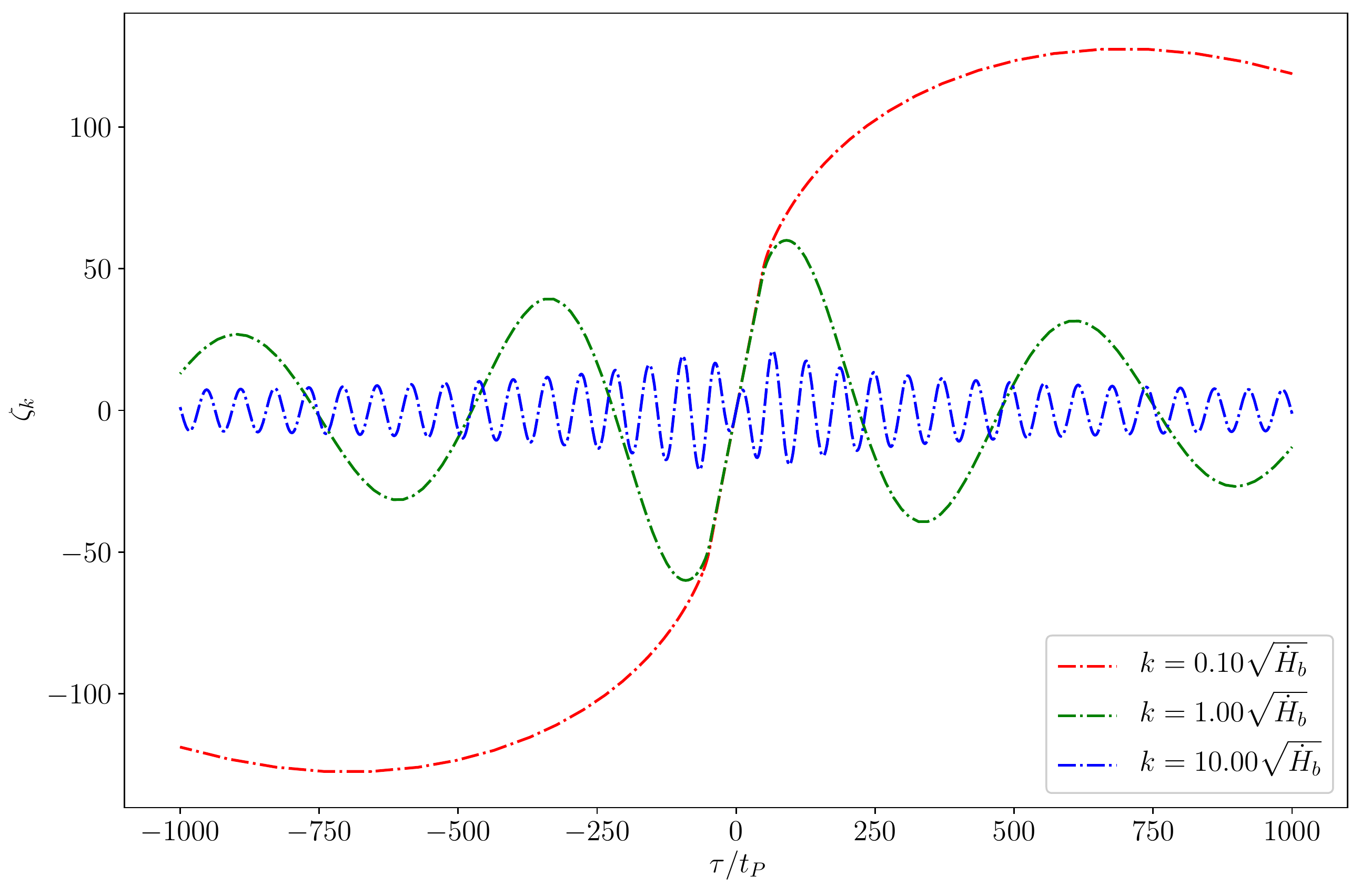}
\caption{\label{fig:scalar_sols} Independent solutions for $\zeta_k$ for three different scales. The figure on the left is with boundary condition set as $\zeta_k(0) = 0$ and $\zeta_k'(0) = 1$ at the bounce, while the figure on the right is with $\zeta_k(0) = 1$ and $\zeta_k'(0) = 0$. Both figures demonstrate that scalar mode solutions are non-singular and stable through the bounce.}
\end{figure}
\begin{align}
    S^{(2)}_{\zeta} =\frac{M_P^2}{2}  \int dt \ d^3 k \ az^2 \left[ \dot{\zeta}_k^2 - \frac{c_s^2 k^2}{a^2} \zeta_k^2 \right], \label{eq:action_curv}
\end{align}
where $c_s$ and $z$ are functions that are both time and scale dependent but reduce to their corresponding standard forms on small scales (large $k$). The exact form of these functions are given by
\begin{align}
    c_s^2 &= \frac{(k/a)^4 H^2 + (k/a)^2 \mathcal{B}_1 + \mathcal{B}_2 }{ (k/a)^4 H^2 + (k/a)^2 \mathcal{A}_1 + \mathcal{A}_2 } \label{eq:cs} \\
    z^2 &= 2 a^2 P \left( \frac{(k/a)^2 +3P }{ (k/a)^2 H^2 + P(3H^2 + P + \dot{H}) } \right), \label{eq:z}
\end{align}
where the following background dependent quantities were introduced to simplify the above expressions:
\begin{align}
    \mathcal{A}_1 &= P (6H^2 + \dot{H} + P) \\
    \mathcal{A}_2 &= 3 P^2 (3H^2 + \dot{H} + P) \\
    \mathcal{B}_1 &= P (12 H^2 + 3 \dot{H} + P) + \dot{H} \left( 2 \dot{H} + \frac{H \ddot{H}}{H} \right) \\
    \mathcal{B}_2 &= P^2 (15 H^2 - P + \dot{H}) - P \dot{H} \left( 12H^2 - 2\dot{H} + \frac{3 H \ddot{H} }{\dot{H}} \right).
\end{align}
These quantities were studied in detail in \cite{Boruah:2017tvg, Boruah:2018pvq} and interested readers are encouraged to refer to them for further discussion. To summarise, it was shown that first of all the sign of the kinetic term is always positive and hence there are no ghost instabilities. Second, that the independent solutions for $\zeta_k$ are stable and non-singular across the bounce both on small and large scales. Figure \eqref{fig:scalar_sols} illustrates this for three sets of these solutions for different scales of $k$. Note that the scales for $k$ are given in terms of $\sqrt{\dot{H}_b} \approx 0.01 M_P $ to give context of the scales in our model. This is by construction since $\dot{H}_b$ depends on the model parameters.

\section{Stability of tensor perturbations through the bounce} \label{tensor}

\begin{figure}[tbp]
\centering 
\includegraphics[width=.49\textwidth]{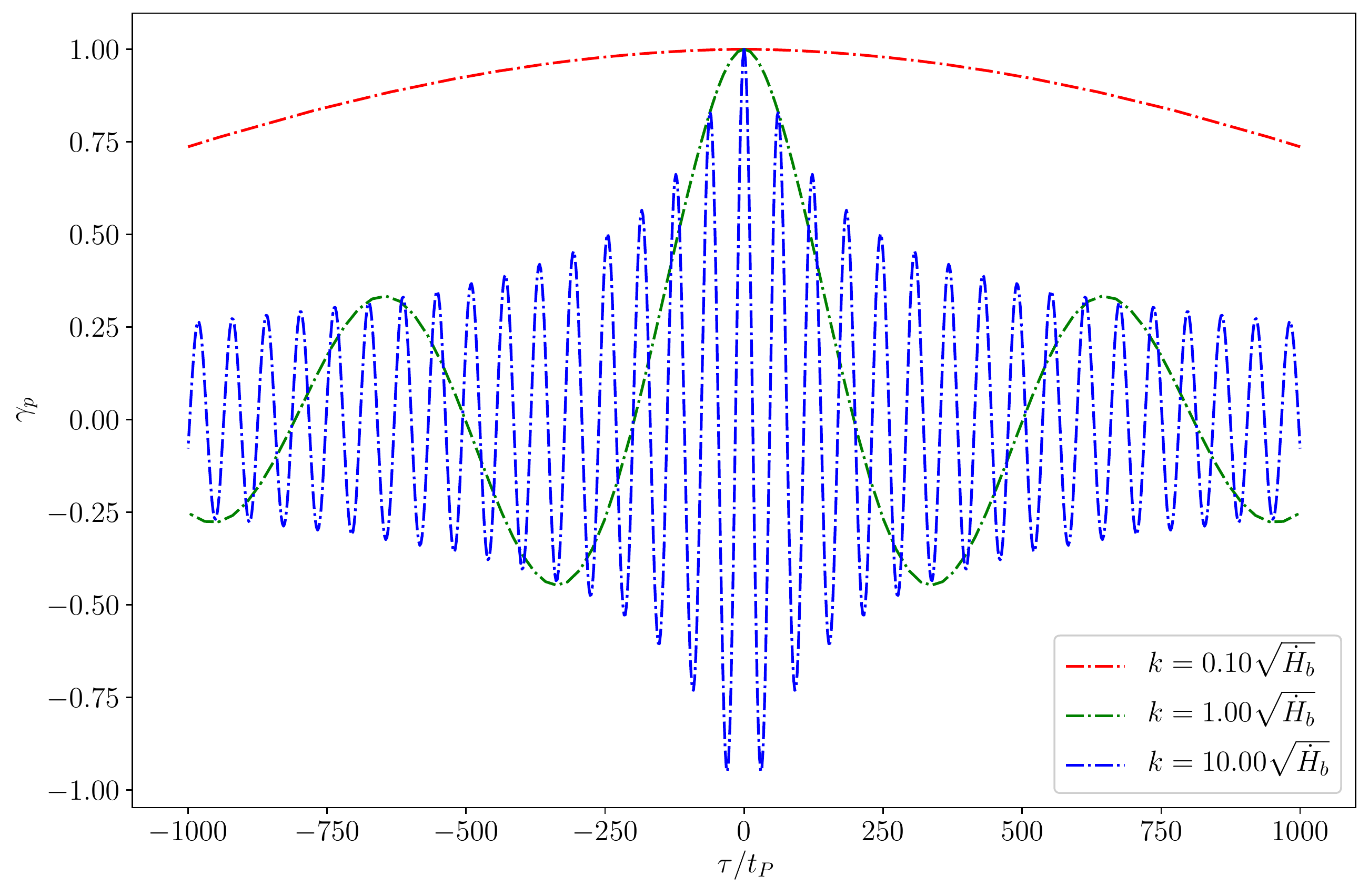}
\hfill
\includegraphics[width=.49\textwidth]{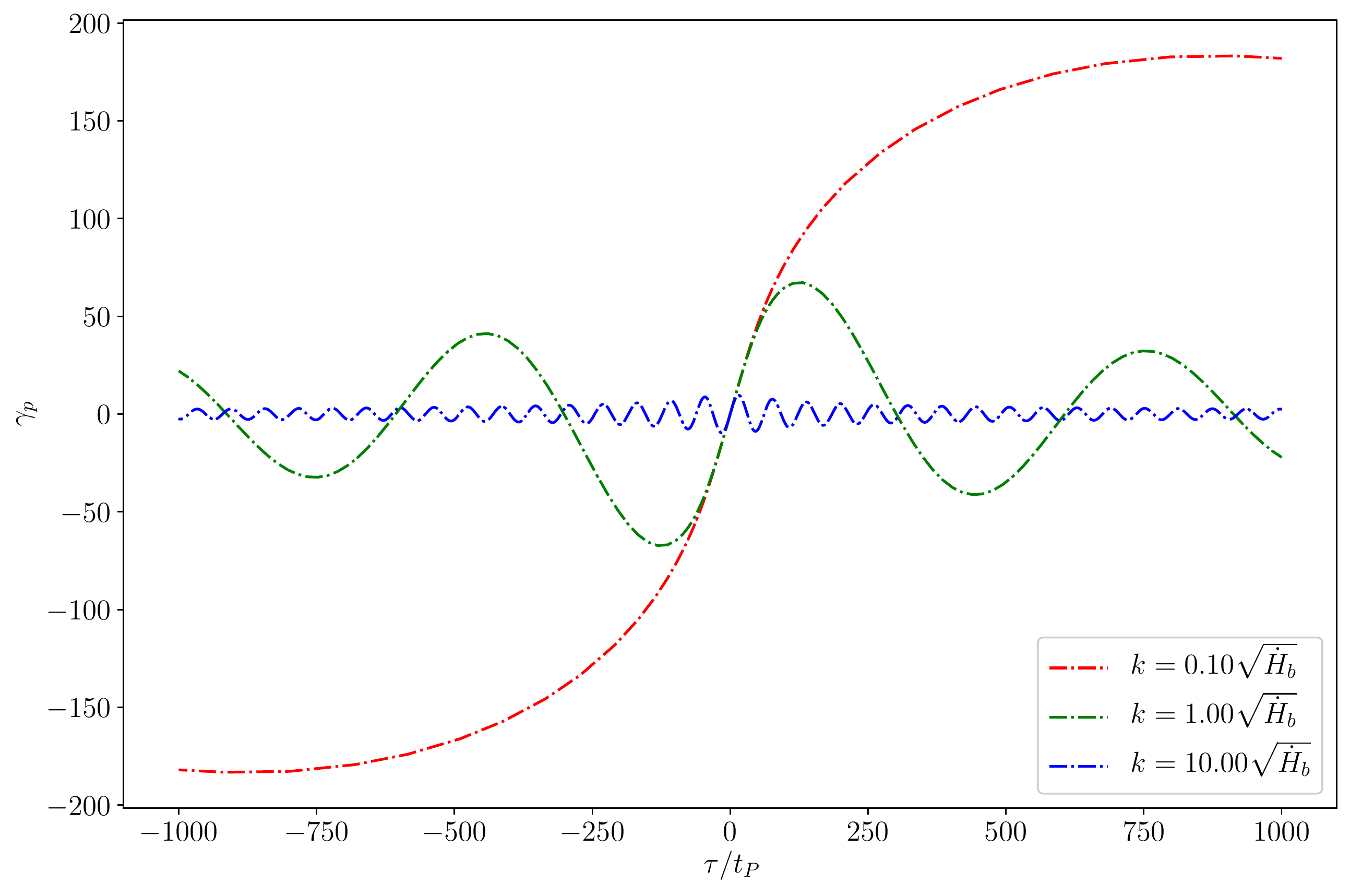}
\caption{\label{fig:tensor_sols} Independent solutions to tensor modes equation of state for three different scales. The figure on the left is with boundary condition set to $\gamma_p(0) = 0$ and $\gamma_p'(0) = 1$, while the figure on the right is with $\gamma_p(0) = 1$ and $\gamma_p'(0) = 0$. Both figures demonstrate solutions are non-singular and stable through the bounce.}
\end{figure}
While the thorough investigations discussed in the last section show that the scalar sector of perturbations in Cuscuton bounce is stable, one could still ask whether the same statement is valid for tensor mode perturbations as well. 
In fact, it is often the case that the analysis of the tensor perturbations is less complicated than the analysis for scalar perturbations. The reason is that since $\sqrt{-g}$ that couples to Cuscuton and matter fields only gets $\zeta$ corrections at second order and terms with spacial covariant derivatives are already at second order in fields variations, the only perturbative contribution to action \eqref{eq:oldaction} from $\gamma_{ij}$ at second order is through the standard Einstein-Hilbert term. Therefore, action for tensor modes in our Cuscuton model is no different than the usual standard,
\begin{align}
    S_\gamma^{(2)} &= \frac{M_P^2}{8} \int d\tau d^3 k  \ a^2 (\gamma_{ij}'^2 - (\nabla \gamma_{ij})^2 ), \label{eq:gw_action}
\end{align}
where the conformal time $\tau$ is defined as $dt = a ~d\tau$ and $\nabla$ is the differential operator for spatial dimensions, so that $(\nabla \gamma_{ij})^2 = \partial_a \gamma_{ij} \partial^a \gamma_{ij} $. The tensor perturbation can be split up into the appropriate tensor polarizations in the $+$ and $\times$ directions,
\begin{align}
    \gamma_{ij} = \sum_{p = +,\times} \gamma_p e^p_{ij} ,
\end{align}
where $\gamma_p$ represents the distinct amplitudes for the two polarization modes of the gravitational waves $p = \times$ and $ p = +$, and $e^p_{ij}$ represent the fixed polarization basis vectors with the property $e^p_{ij} e^{ij}_{p'} = 2\delta^p_{p'}$. Without loss of generality, assuming that the propagation direction for the gravitational waves is in the $z$-direction, the tensor perturbation can be written in matrix representation as
\begin{align}
    \gamma_{ij} = 
    \begin{pmatrix}
        \gamma_+ & \gamma_\times & 0  \\
        \gamma_\times & - \gamma_+ & 0 \\
        0 & 0 & 0 
    \end{pmatrix}.
\end{align}
Then converting to Fourier space, the action in eq. \eqref{eq:gw_action} simplifies to 
\begin{align}\label{eq:gw_action2}
    S_\gamma^{(2)} = \frac{M_P^2}{4} \sum_{p = \times, +} \int d\tau d^3k \ a^2  ( \gamma_p'^{2}  - (\nabla \gamma_p)^2),
\end{align}
Next, similar to scalar modes, one can obtain the equation of motion for $\gamma_p$ and then check to see if the independent solutions for tensor modes are non-singular and stable. Once again we see that the solutions are healthy for different scales in figure \eqref{fig:tensor_sols} for a set of arbitrary conditions injected at the bounce. Therefore, there is no instability associated with tensor perturbations in Cuscuton bounce either.  

\section{Power spectrum for scalar modes in single field Cuscuton bounce}\label{PSsingle}

\begin{figure}[tbp]
\centering 
\includegraphics[width=.49\textwidth]{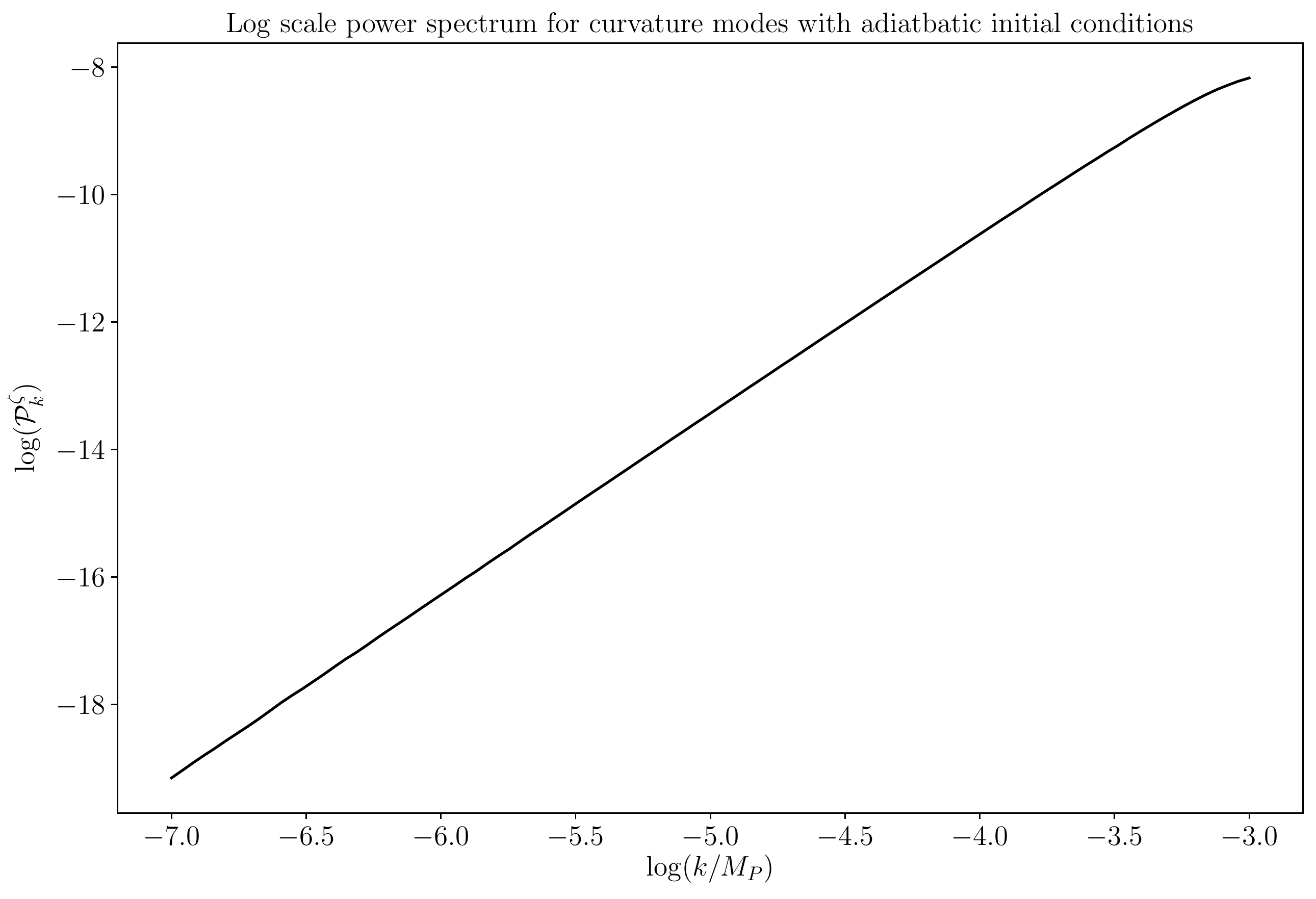}
\hfill
\includegraphics[width=.49\textwidth]{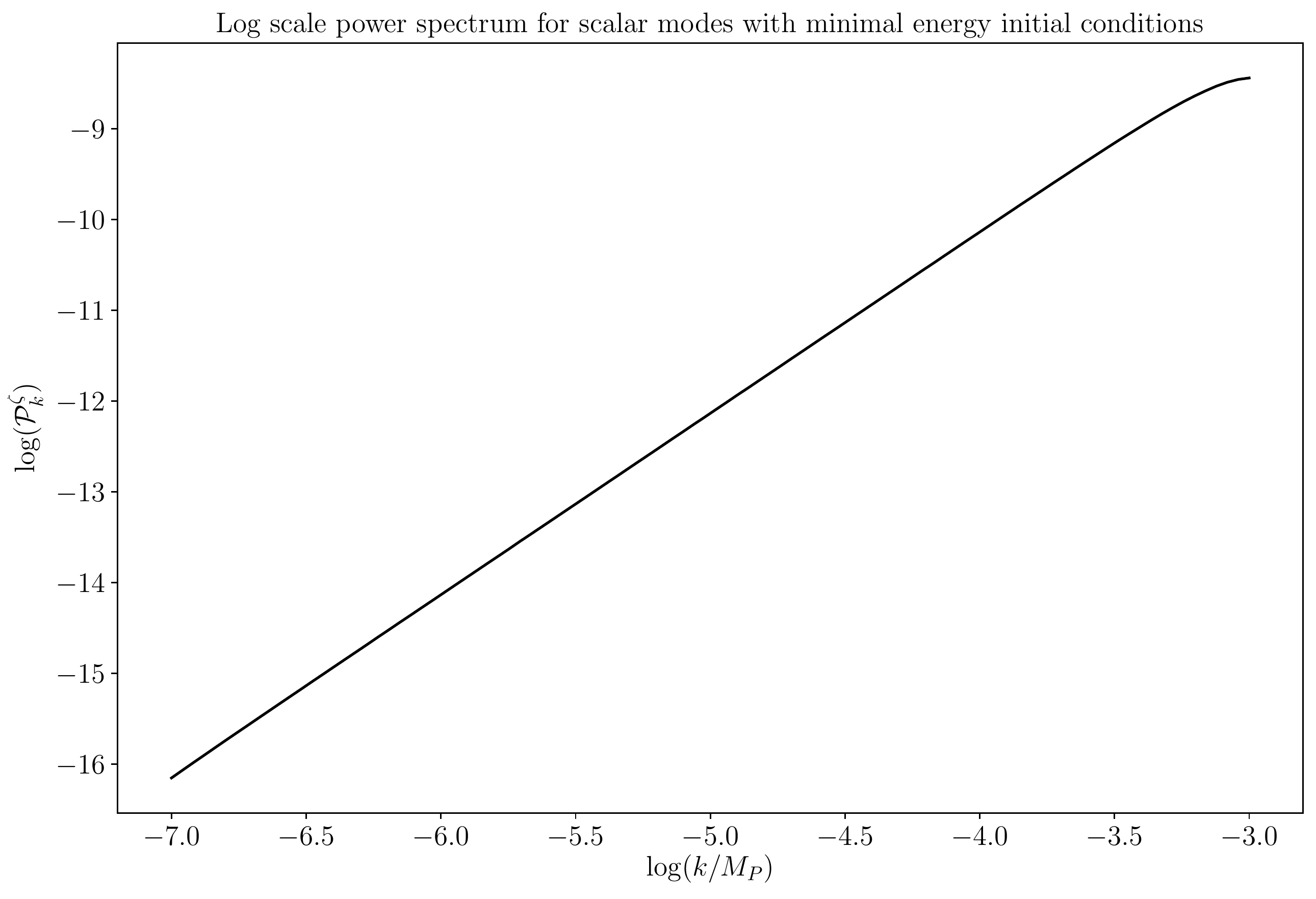}
\caption{\label{fig:ps_scalar}Log scale power spectrum for scalar perturbations (left) imposing adiabatic initial conditions at $\tau_i = -10^{8} t_P$ and (right) imposing instantaneous minimal energy conditions at the bounce evaluated at a post time $\tau_f = 1000 t_P$. Both cases exhibit very blue power spectra.}
\end{figure}

As discussed in Section \eqref{intro}, while existence of an stable bounce by itself has interesting implications for early universe and big bang singularity, another intriguing question is whether it could also provide an alternative for inflation. In particular, if it can provide a mechanism to produce curvature perturbations consistent with current data from adiabatic vacuum quantum fluctuations or other initial/boundary conditions. In this section we investigate this question for single field Cuscuton bounce and argue that this scenario as it stands, cannot produce a near scale-invariant scalar power spectrum. 

To study the generation of scalar perturbations, we start by introducing the Mukhanov-Sasaki variable
\begin{align}
    v_k = M_P ~z(\tau,k) \zeta_k,
\end{align}
and turning the action \eqref{eq:action_curv} into the canonical form,
\begin{align}
   S^{(2)}_{v_k} =\frac{1}{2}  \int d\tau \ d^3 k \ \left[ v_k'^2 + \left( \frac{z''}{z} - c_s^2 k^2 \right) v_k^2 \right].
\end{align}
This enables us to apply the standard field theory quantization scheme, where $v_k$ is promoted to operators $\hat{v_k}$. We refer readers to \cite{Mukhanov:2007zz} for a review on quantizing cosmological perturbations in curved space-times. From here on we reserve the notation $v_k$ to denote the mode functions corresponding to that operator, which also satisfy the classical equation of state, 
\begin{align}
    v_k'' + \left( c_s^2 k^2 - \frac{z''}{z} \right) v_k = 0 \label{eq:scalar_eom}.
\end{align}
In order to calculate the amplitude of perturbations one needs to a make choice about initial/boundary conditions as well. We investigated the solutions to \eqref{eq:scalar_eom} under three different possibilities and showed that neither of them produce a near scale-invariant scalar power-spectrum.

First, we started by setting the initial conditions at infinite past to adiabatic vacuum state. Like inflation, one could argue that generating all the structure in the universe out of vacuum quantum fluctuation is too impressive of an idea not to pursue. However, as we know, since the cosmological background is time dependent, the state of minimum energy also changes in time. In adiabatic regimes, where WKB approximations is satisfied, the adiabatic vacuum initial condition \cite{Birrell:1982ix} remains close to minimum energy state. In practice, to impose that condition numerically, we selected the initial time, $\tau_i$, long before the bounce such that the condition of $ k^2 \gg \frac{z''}{z c_s^2 }$ was satisfied and then imposed the following relations 
\begin{align}
    v_{k}(\tau_i) = \frac{1}{\sqrt{2 \omega_{S}}} e^{-i \omega_{S} \tau_i}, \qquad
    v_{k}'(\tau_i) = - i \sqrt{\frac{\omega_{S} }{2}} e^{- i \omega_{S} \tau_i}, \label{eq:adiabatic}
\end{align}
where we have defined 
\begin{align}
 \omega_{S}^2\equiv c_s^2 k^2 - \frac{z''}{z}. 
 \end{align}
The second possibility we considered was imposing instantaneous minimal energy condition at the bounce, $\tau_b = 0$. Motivation for this choice is that since the model is symmetric in time around the bounce, an underlying fundamental symmetry may enforce the fluctuation into ground state at $\tau_b = 0$, in order to preserve the symmetry\footnote{Also see \cite{Boyle:2018tzc} for a different proposal regarding a CPT symmetric universe at big bang.}. This condition was set by imposing, 
\begin{align}
    v_{k}(\tau_b) = \frac{1}{\sqrt{2 \omega_{S}}},  \qquad 
    v_{k}'(\tau_b) = -i \sqrt{\frac{\omega_{S} }{2}}.
\end{align}
For either of the conditions listed above, we solved the equation of motion \eqref{eq:scalar_eom} and estimated the dimensionless power spectrum for $\zeta$ at a post bounce time through 
\begin{align}
    \mathcal{P}_k^{\zeta_k} (\tau_f) = \frac{k^3}{2\pi^2} | \zeta_k (\tau_f) |^2 = \frac{k^3}{2 \pi^2 M_p^2} \frac{|v_k(\tau_f)|^2}{ z^2(\tau_f)}.
\end{align}
Figure \eqref{fig:ps_scalar} demonstrates the logarithmic scale dependence of the scalar power spectrum against wave number $k$ for the two different initial condition hypotheses. In both cases we see that scale-invariance is not achieved, and both spectra are strongly blue.
\begin{figure}[tbp]
\centering 
\includegraphics[width=0.9\textwidth]{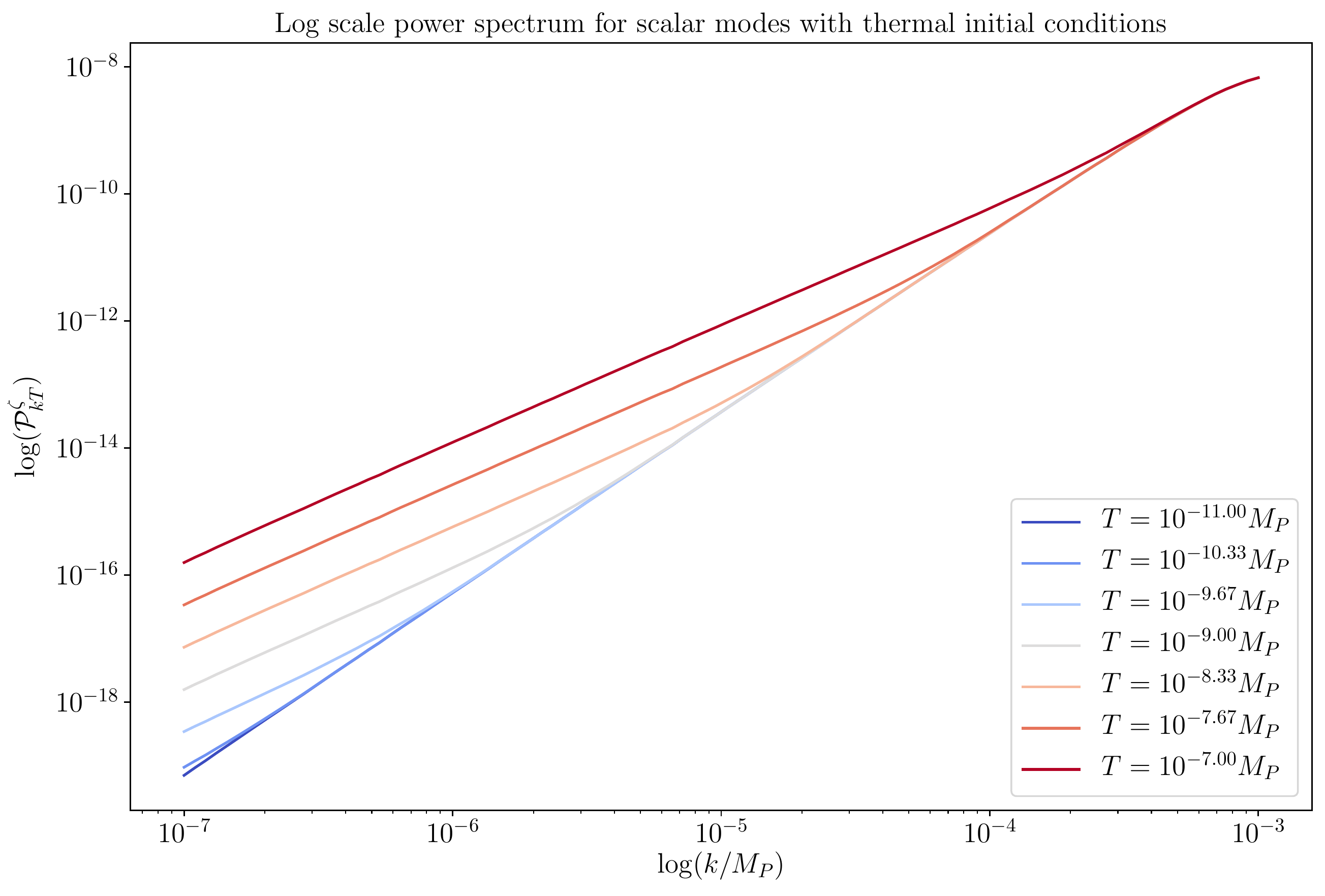}
\caption{\label{fig:thermal_ps} The power spectrum for $\zeta$ in the case of thermal initial conditions. Again the initial conditions were injected before the bounce at $\tau_i = -10^{8} t_P$ and evaluated after the bounce at $\tau_f = 1000 t_P$. In this case the temperature of the background determines the tilt of the power spectrum, but is still unable to acquire scale-invariance.}
\end{figure}

In addition to these two initial conditions, we also considered thermal initial conditions \cite{Magueijo:2002pg, Ferreira:2007cb, Magueijo:2008pm, Agarwal:2014ona}. If universe existed for a long period of time before the bounce and there were additional interaction channels for fluctuations, they could have settled into a thermal equilibrium as well. In this case, we assumed the thermal energy density in fluctuations is subdominant to the background energy density in order to ignore back reaction effects. If we assume thermal initial conditions, the resulting vacuum power spectrum will be adjusted by an additional factor,
\begin{align}
    \mathcal{P}^{\zeta_k}_{\text{ther. ini.}} = (1 + 2 \langle n_k \rangle_{\text{ini}}) \mathcal{P}^{\zeta_k}_{\text{vac. ini.}},
\end{align}
where $\langle n_k \rangle$ is the standard Bose-Einstein particle occupation number,
\begin{align}
    \langle n_k \rangle_{\text{ini}} = \frac{1}{ e^{\frac{c_sk}{aT}} - 1 }.
\end{align}
Note that $c_s, a$ and $T$ are evaluated at the initial time, assuming that scalar mode excitations were thermalized with temperature $T$.\footnote{Note again that in our calculations, we are setting $a=1$ at the bounce. The comoving wave number $k$ represents the physical scale at the bounce and not the physical scale at present time.} Therefore, taking thermal initial condition, the power spectrum plotted on the left in figure \eqref{fig:ps_scalar} is adjusted by this new factor, which is demonstrated in figure \eqref{fig:thermal_ps} for different values of temperature $T$.

We can see all three sets of initial conditions for the scalar modes result in power spectra that exhibit a strongly blue tilt, inconsistent with observations. This is strong evidence that a Cuscuton bounce using only one matter field is not enough to produce a scale-invariant power spectrum for adiabatic perturbations, unless some new mechanism is introduced. We end the discussion for scalar modes in a single field Cuscuton bounce here. Since the modification to Cuscuton in section \eqref{twofield} does not affect tensor modes, we will return to tensor modes in single field Cuscuton bounce in section \eqref{PStensor}.

\section{Two field Cuscuton bounce and power spectrum for entropy perturbations}\label{twofield}

As we discussed in previous chapter, the power spectrum for curvature perturbations in a single field Cuscuton bounce is strongly blue. In fact, this result is consistent with what has been observed in other generic bounce models proposed in the past. For example, in the Pre-Big Bang scenario \cite{Gasperini:1992em, Gasperini:2002bn}, inspired by superstring theory, the power spectrum for curvature perturbations exhibits a strong blue tilt with $n_s = 4$. It has also been shown that the single field Ekpyrotic models \cite{Khoury:2001wf}, generically produce a blue power spectrum with a tilt of $n_s = 3$ \cite{Creminelli:2004jg, Tseng:2012qd, Battarra:2013cha}. As we know, this is contradictory to the observational precision measurements of the scalar power spectrum, such as those obtained through CMB measurements \cite{Aghanim:2018eyx}. However, note that mathematically speaking, scalar modes can also include entropy (or isocurvature) perturbations in addition to the curvature (adiabatic) perturbations. While the observations put very tight constraints on the present day amplitude of entropy perturbations contributing to scalar anisotropies \cite{Akrami:2018odb}, they still leave the possibility that given the numerous degrees of freedoms present in early universe, these entropy modes were generated but then converted to curvature perturbations. 
In this section we will provide an example of how entropy perturbations with a nearly scale-invariant power spectrum could have been generated in Cuscuton bounce. This process is sometimes referred to as the entropic mechanism \cite{Buchbinder:2007ad, Notari:2002yc}, in which an additional field in the model sources the entropy perturbations that would later be converted into curvature perturbations \cite{Lehners:2007ac, Ijjas:2014fja, Li:2013hga}.\footnote{Another possible mechanism inspired by S-branes in string theory, that has been more recently suggested in \cite{Brandenberger:2020tcr, Brandenberger:2020eyf, Brandenberger:2020wha} could also lead to near scale-invariant power spectrum. The key difference between the two mechanisms is that as opposed to adding a new degree of freedom in entropic mechanism, in S-brane scenario, a delta-function potential gets added to effective potential at the bounce, and is the driving force for the actual bounce.
Since our Cuscuton model gives us a non-divergent solution all across the bounce without needing an additional mechanism to generate the bounce itself, we explore the implications of the entropic mechanism for our power spectrum.}

We start by adding a second field, $\chi$, to our action \eqref{eq:oldaction}. Once again we choose the simplest case such that the additional field only has a stabilized kinetic term ($\dot{\chi}_0 = 0$), also subdominant to the background. This field will not contribute to the Friedmann equations at zeroth order, and leaves the background dynamics unchanged. Since the field is massless, we need to allow for non-minimal coupling between this field and the dominant matter field to produce an effective mass and to source perturbations in $\chi$. In fact, as we see below, if this coupling $F(\dot{\pi},\nabla_i \pi,\dots)$ is proportional to the dominant background matter density, $\rho_m$, then power spectrum will automatically be scale-invariant.  
In this case, the action is given by 
\begin{align} \label{actionchi}
    S = \int d^4 x \sqrt{-g} \left[ \frac{M_P^2}{2} R - \frac{1}{2} D_\mu \pi D^\mu \pi - \frac{1}{2} F(\dot{\pi},\nabla_i \pi,\dots ) D_\mu \chi D^\mu \chi - \mu^2 \sqrt{-D_\mu \varphi D^\mu \varphi} - V(\varphi)  \right],
\end{align}
where $\varphi$ is the Cuscuton field, $\pi$ is the scalar field from before, and $\chi$ is the new additional scalar field.\footnote{One difference between the Ekpyrotic and Cuscuton entropic mechanisms is that in the Ekpyrotic scenario, the dynamics are generated from the Ekpyrotic potential for the matter field \cite{Khoury:2001wf, Buchbinder:2007ad}, while in the Cuscuton scenario, both potentials for the scalar fields are zero and the dynamics are generated from the Cuscuton potential.}

As before, variation of the action at zeroth order in perturbations $\pi=\pi_0(t)+\delta\pi (x,t)$ and $\chi=\chi_0(t) + \delta \tilde{\chi}(x,t)$ around the FLRW metric leads to background equations of motion. The equation of motion for the Cuscuton field remains unchanged compared to the single field case, while for the scalar fields $\pi_0$ and $\chi_0$ we have
\begin{align}
    - \frac{1}{2} \dot{\chi_0}^2 F'(\dot{\pi}_0) + 3 H \dot{\pi}_0 + \Ddot{\pi}_0  &= 0 \label{eq:premod}  \\
   3 F(\dot{\pi}_0) H \dot{\chi}_0 + F(\dot{\pi}_0) \Ddot{\chi}_0 + \ddot{\pi}_0 \dot{\chi}_0 F'(\dot{\pi}_0) &= 0.
\end{align}
Since we are assuming the energy density of the second field to be subdominant to the preexisting matter field and negligible for the background, this condition is satisfied if we simply set $\dot{\chi_0} = 0$ which in return makes the $\pi$-field  and the Cuscuton field resume their background evolution obtained in previous section. We will show later that this solution is indeed stable and $\delta \tilde{\chi}$ do not exhibit any instabilities. Therefore our assumption is self-consistent. To summarize, under this simplification, adding a non-minimally coupled stabilized scalar field to our Cuscuton bounce model does not change the background dynamics. 

\begin{figure}[tbp]
\centering 
\includegraphics[width=0.9\textwidth]{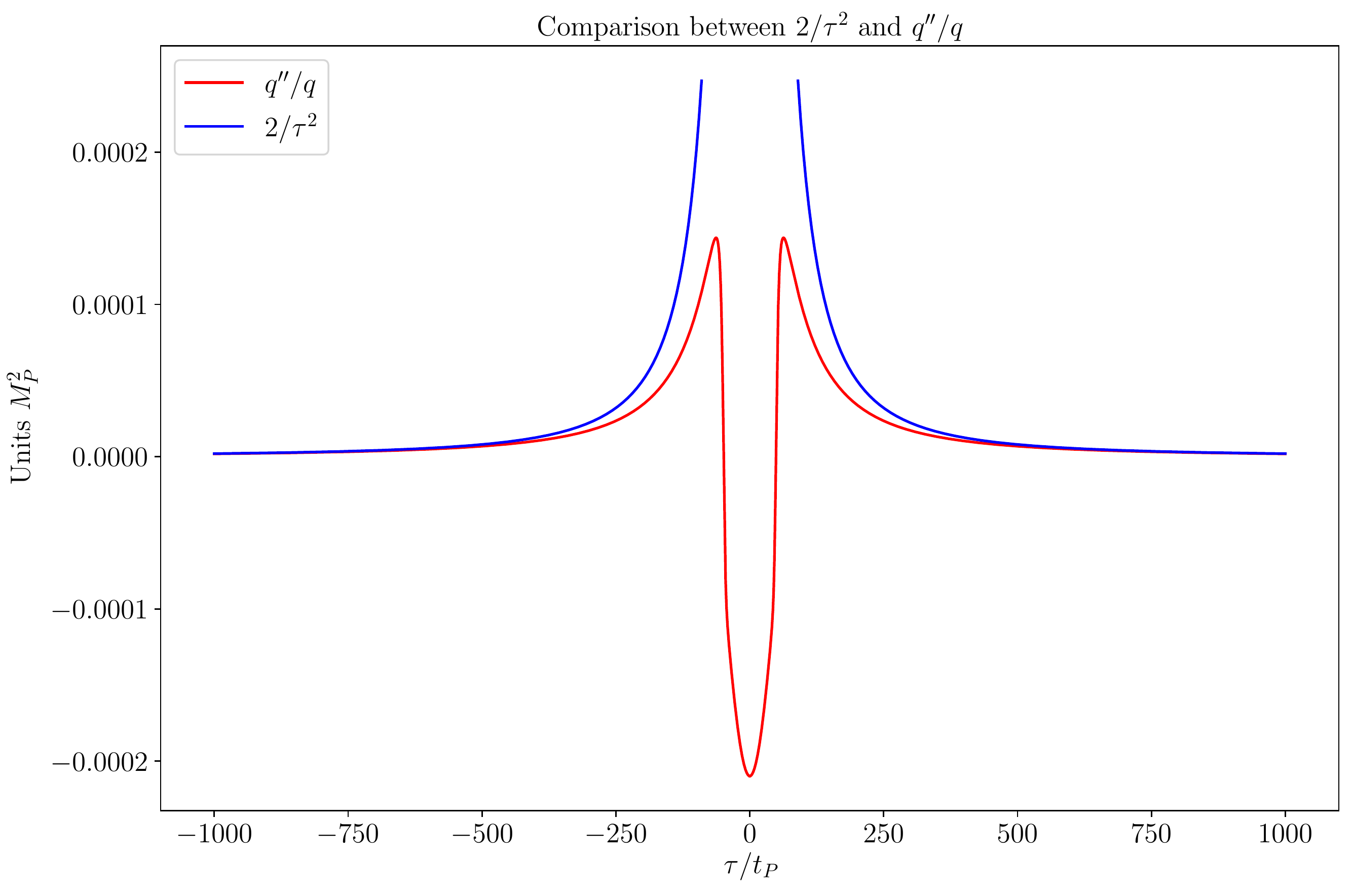}
\caption{\label{fig:qppbyq_vs_tau2} Comparison between the scale-invariance case of $2/\tau^2$ (blue) and the Cuscuton $q''/q$ (red). By construction, $q''/q$ approaches $2/\tau^2$ far from the bounce.}
\end{figure}

Next, we study the behaviour of the cosmological perturbations under this additional coupling in the model. In all our numerical calculations we are working in Planck units, still in order to keep track of dimensions if we take $\chi$ to have dimension of mass, then $F(\dot{\pi}_0)$ is dimensionless. Therefore, from here on what we refer to as entropy perturbations is described by the dimension-less variable $\delta \chi \equiv \frac{\delta \tilde{\chi}}{M_p }$. Substituting this back in \eqref{actionchi} and calculating the contribution to the second order action from $\chi$-field yields
\begin{align}
    S_{\chi}^{(2)} &= \frac{M_p^2}{2} \int d\tau d^3 x a^2 F(\dot{\pi}_0) \left[ (\delta \chi')^2 - (\partial \delta \chi)^2 \right] = \frac{M_p^2}{2} \int d\tau d^3 x~ q^2 \left[ (\delta \chi')^2 - (\partial \delta \chi)^2 \right], \label{eq:chiaction1}
\end{align}
where we have also defined $q^2 \equiv a^2 F(\dot{\pi}_0)$. Introducing the canonical variable $u$,
\begin{align}
    u  =M_p ~q ~\delta \chi,
\end{align}
the action \eqref{eq:chiaction1} can be rewritten as
\begin{align}
    S_{\chi}^{(2)} &= \frac{1}{2} \int d\tau d^3 x \left[ u'^2 - (\partial u)^2 + \frac{q''}{q} u^2 \right].
\end{align}
The equation of motion in Fourier space is then given by
\begin{align}\label{modequ}
    u_k'' + \omega_u^2 (\tau, k) u_k = 0,
\end{align}
where the effective frequency $\omega_u$ for these modes is
\begin{align}\label{modequ2}
    \omega_u^2 (\tau,k) = \left( k^2 - \frac{q''}{q} \right).
\end{align}
Up to this point, we have not yet provided any description on the analytical dependency of the coupling function $F(\dot{\pi})$ to $\dot{\pi}$. However, a very important lesson familiar to most early universe cosmologists is that if 
\begin{align}\label{qeq}
    \frac{q''}{q} \sim \frac{2}{\tau^2},
\end{align}
then \eqref{modequ} turns into a modified Bessel equation, which upon imposing vacuum initial conditions leads to nearly scale-invariant power spectrum. This is the magic that occurs to spectator fields on a de Sitter space-time back grounds or tensor/scalar modes during slow-roll inflation which also effectively behave as spectator fields on a quasi-de Sitter background. Note that in general, equation \eqref{qeq} has two independent solutions:
\begin{align}
    q(\tau) = \frac{1}{\Lambda\tau} +M^2 \tau^2 ,
\end{align}
and different values of constant $\Lambda$ and $M$ could in principle correspond to different mass scales \cite{Khoury:2008wj, Geshnizjani:2011dk}. For example, in the case of a slow-roll inflationary model, for scalar modes $q\sim a\sqrt{\epsilon} \sim \frac {\sqrt{ \epsilon}}{H\tau}$ approximately and for tensor modes $q\sim a \sim \frac {1}{H\tau}$. Therefore, in this case $M=0 $ and the value of $\Lambda\sim \frac {H} {\sqrt{ \epsilon}}$ for scalars and $\Lambda\sim H$ for tensors determines the amplitude for perturbations. Similarly in our model, we can also find a general form of the coupling function $F(\dot{\pi}_0)$ such that entropy modes are nearly scale-invariant. 
In principle we could allow for the most generic case, where both $\Lambda$ and $M$ contributions exist. However, since $\tau=0$ corresponds to the time of the bounce, unless $\Lambda$ contribution is strictly zero then $\frac {1}{\Lambda\tau} $ will always dominate the behaviour of $q$. Therefore, we proceed by considering $q(\tau) \approx \frac{1}{\Lambda \tau}$ and obtain the dependence of $F(\dot{\pi})$ to $\tau$, 
\begin{align}\label{Ftau}
 F(\dot{\pi}_0)\approx \frac{1}{a^2} \left( \frac{1}{\Lambda\tau} \right)^2. 
\end{align}
Note that at this point, we are simply focusing on solutions that can lead to scale-invariance itself. However, the next order effects which describe the precision value of spectral index have to be determined by accuracy of the relation \eqref{Ftau}, which in the language of inflation is translated to slow variation of $\Lambda(\tau)$ i.e. $H(\tau)$ in time, due to slow-roll parameters. 
\begin{figure}[tbp]
\centering 
\includegraphics[width=.49\textwidth]{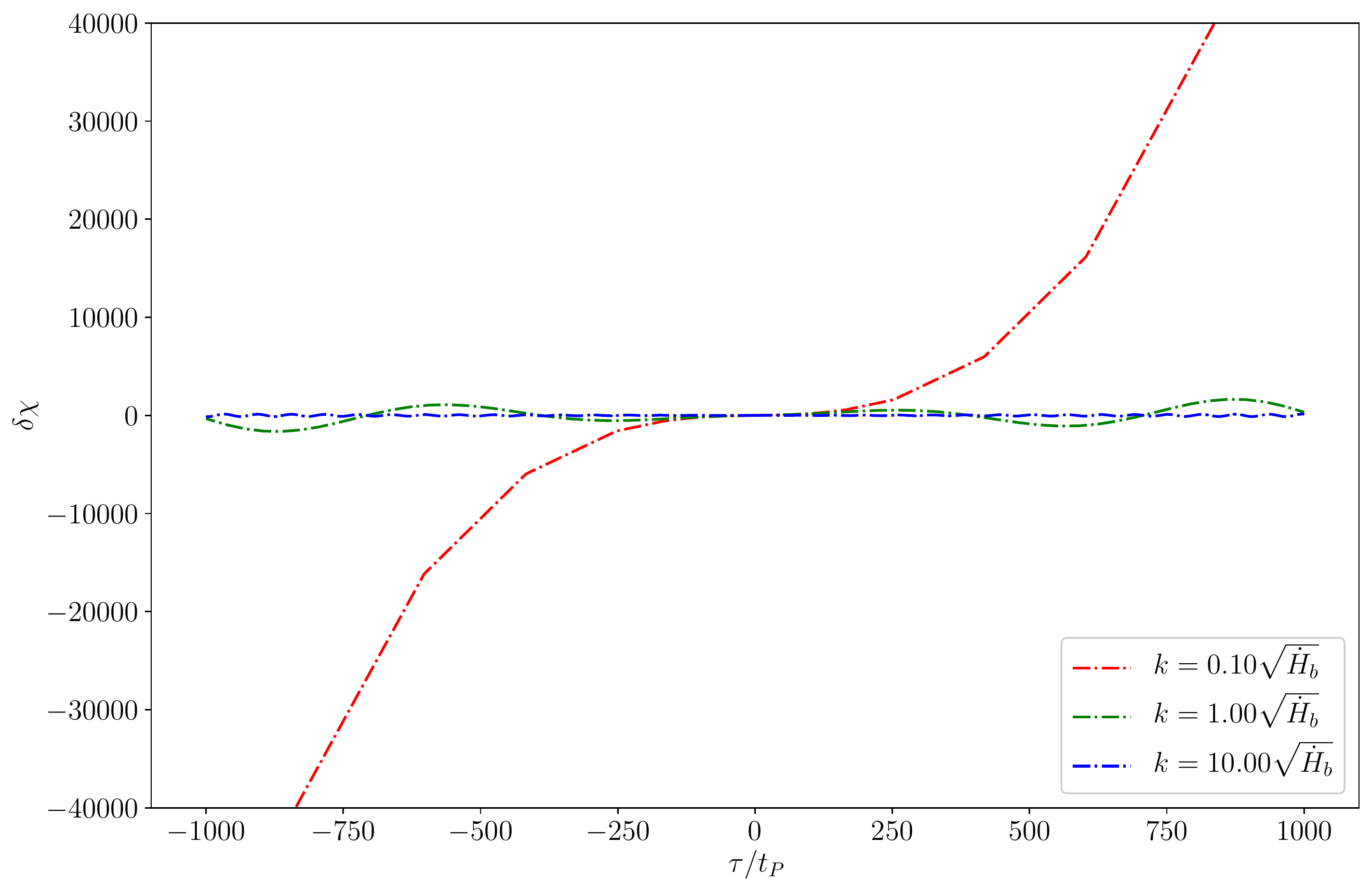}
\hfill
\includegraphics[width=.49\textwidth]{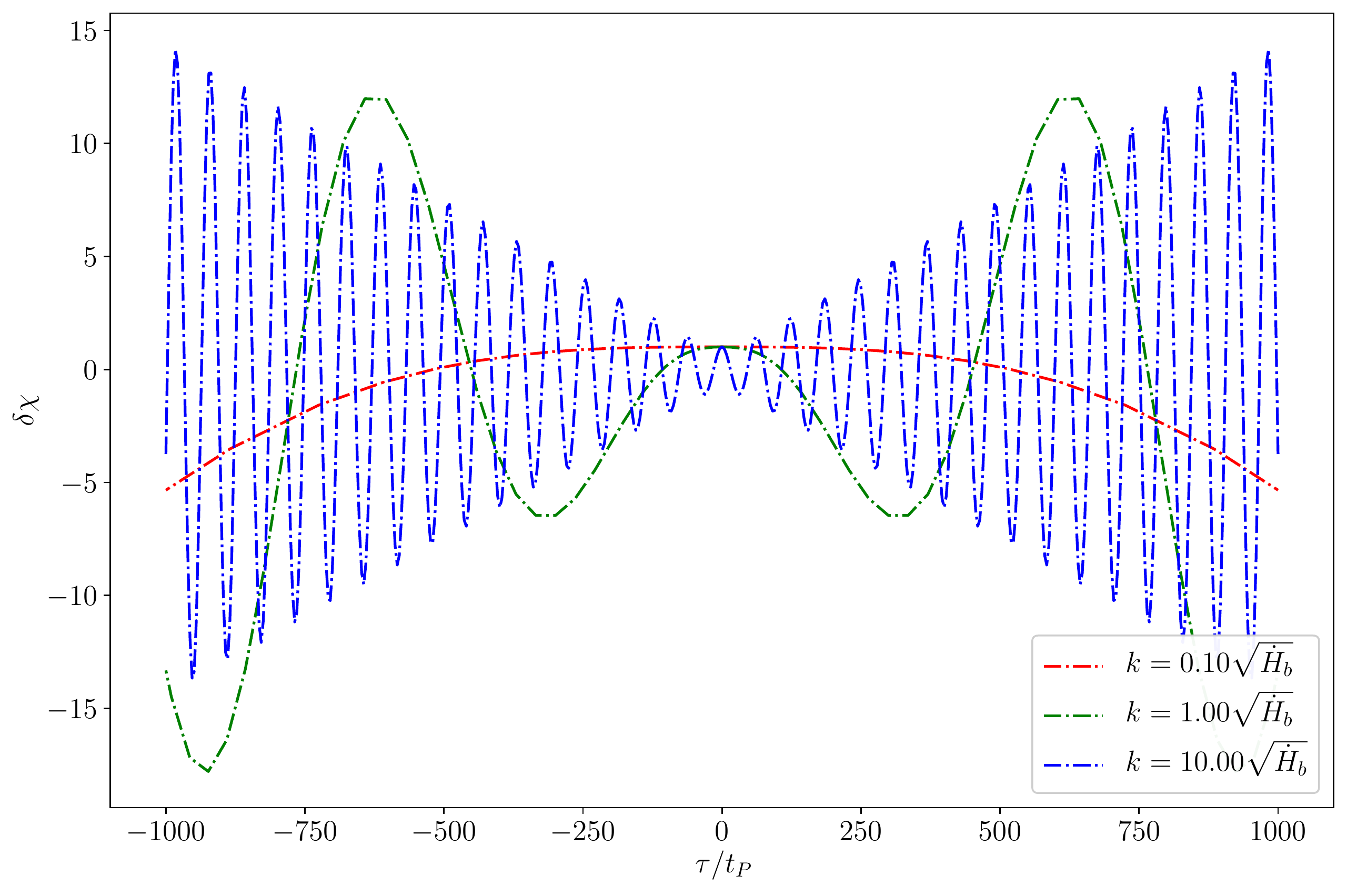}
\caption{\label{fig:deltachi} Two independent solutions for mode equation of $\delta \chi$ perturbations. The figure on the left corresponds to  $\delta \chi(0) = 0$ and $\delta \chi'(0) = 1$, while the figure on the right is obtained by setting $\delta \chi(0) = 1$ and $\delta \chi'(0) = 0$. Both figures admit non-singular solutions through the bounce.}
\end{figure}

 Next taking into account that the bounce transition period, where Cuscuton modifications become significant for the background, is a very brief period in Planck units (around $250 ~t_p$, see figure \eqref{fig:background}) and that it is unlikely to have a significant impact on the power spectrum on cosmological scales, we postulated that it should be sufficient that relation \eqref{Ftau} be satisfied when $\rho_m$ is dominant. However, in that regime $a^2 \propto \tau$ and $\rho_m \propto a^{-6}\propto \tau^{-3}$ which implies $\frac{1}{a^2\tau^2} \propto \rho_m $ and we can consider 
\begin{align}\label{Frho}
    F(\dot{\pi}_0)\equiv \frac{ 1}{\Lambda^2M_p^2}  \left (\frac 12 \dot{\pi}^2_0 \right ) \approx \left (\frac { H^2} {\Lambda^2 }\right)_{a~ \gg ~a_b }. 
\end{align}
In other words, interestingly in order for the theory to produce a near scale-invariant power spectrum, $\chi$ field needs to have the most natural non-minimal coupling to $\pi$, i.e. its energy density.
Expressing this relation in terms of the variable $q$, yields
\begin{align}\label{qdef}
    q(\tau) = \frac{ 1}{\Lambda M_p}\sqrt{\frac 12 \dot{\pi}^2_0}.
\end{align}
Figure \eqref{fig:qppbyq_vs_tau2} shows the comparison between $\frac{q''}{q} \approx \frac{2}{\tau^2}$ \footnote{Which is the approximate time dependence of $q''/q$ in slow-roll inflationary models.} and the Cuscuton model obtained from \eqref{qdef}. As expected, except in the vicinity of the bounce, the coupling function in \eqref{Frho} leads to $\frac{q''}{q}$ tracking $\frac{2}{\tau^2}$ closely.

The magnitude of the coupling constant $\Lambda$ adjusts the amplitude of the power spectrum and it can be determined by comparison to the observational constraints. Physically, it represents the strength of the non-minimal coupling of $\chi$-fields to the density of $\pi$ which could arise from the underlying fundamental theory. 

Once we assert the form of the coupling function from \eqref{Frho}, then substituting $q(\tau)$ from \eqref{qdef} into \eqref{modequ2}, the mode equation \eqref{modequ} can be solved numerically. Again the advantage of our model is that since $\frac{q''}{q}$ does not exhibit any divergences (see Figure \eqref{fig:qppbyq_vs_tau2}), the differential equation is non-singular at all times and does not require any {\it ad hoc} matching conditions.   
\begin{figure}[tbp]
\centering 
\includegraphics[width=0.9\textwidth]{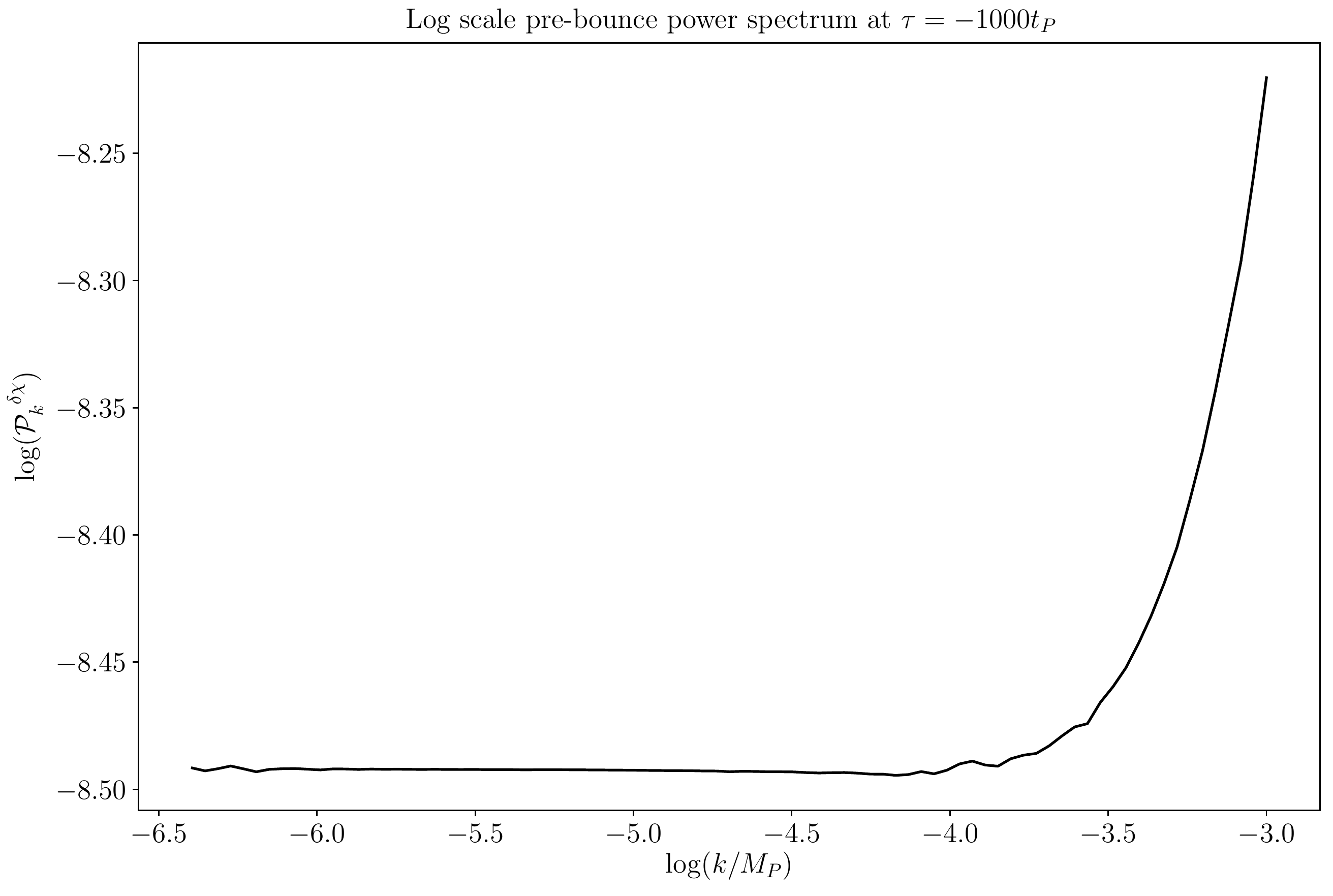}
\caption{\label{fig:pre_bounce_ps_final}Pre-bounce power spectrum for $\chi$-field perturbations evaluated at final time $\tau_f = -1000 t_P$ before Cuscuton modifications start becoming significant. The adiabatic vacuum initial condition is set at $\tau_i = -10^{-8} t_P$ and with $\Lambda = 10^{-3.5}M_p$.}
\end{figure}
As we discussed before, by construction we expect that imposing the adiabatic vacuum initial conditions would lead to a scale-invariant power spectrum. However, to test that, we need to impose the initial conditions numerically similar to the single field model. Therefore, we require 
\begin{align}
    u_k(\tau_i) = \frac{1}{\sqrt{2 \omega_u}} e^{-i \omega_u \tau_i}, \qquad u_k'(\tau_i) = -i \sqrt{\frac{\omega_u}{2}} e^{-i \omega_u \tau_i},
\end{align}
at an initial time $\tau_i$, such that $k |\tau_i| \gg 1$. Then, we estimate the power spectrum for entropy perturbations at some later ``snapshot'' time, $\tau_f$, before and after the bounce by
\begin{align}
    \mathcal{P}^{ \delta\chi}_k (\tau_f) = \frac{k^3}{2\pi^2} | \delta \chi_k (\tau_f) |^2 = \frac{k^3}{2 \pi^2} \frac{|u_k (\tau_f)|^2}{M_p^2 q(\tau_f)^2}.
\end{align}
 Figure \eqref{fig:pre_bounce_ps_final} displays an example of a pre-bounce power spectrum, where we have set $\Lambda = 10^{-3.5}M_p$ and $\tau_f = -1000 t_P$. The resulting power spectrum is indeed scale-invariant except for values of $k$ that correspond to modes that do not cross the freezing horizon i.e. $k^2\gtrsim \frac{q''}{q}|_{t_f}$ and remain oscillatory. In our model these modes are close to Planck scale at the bounce ($k\gtrsim 10^{-4} M_P$). 
\begin{figure}[tbp]
\centering
\includegraphics[width=0.85\textwidth]{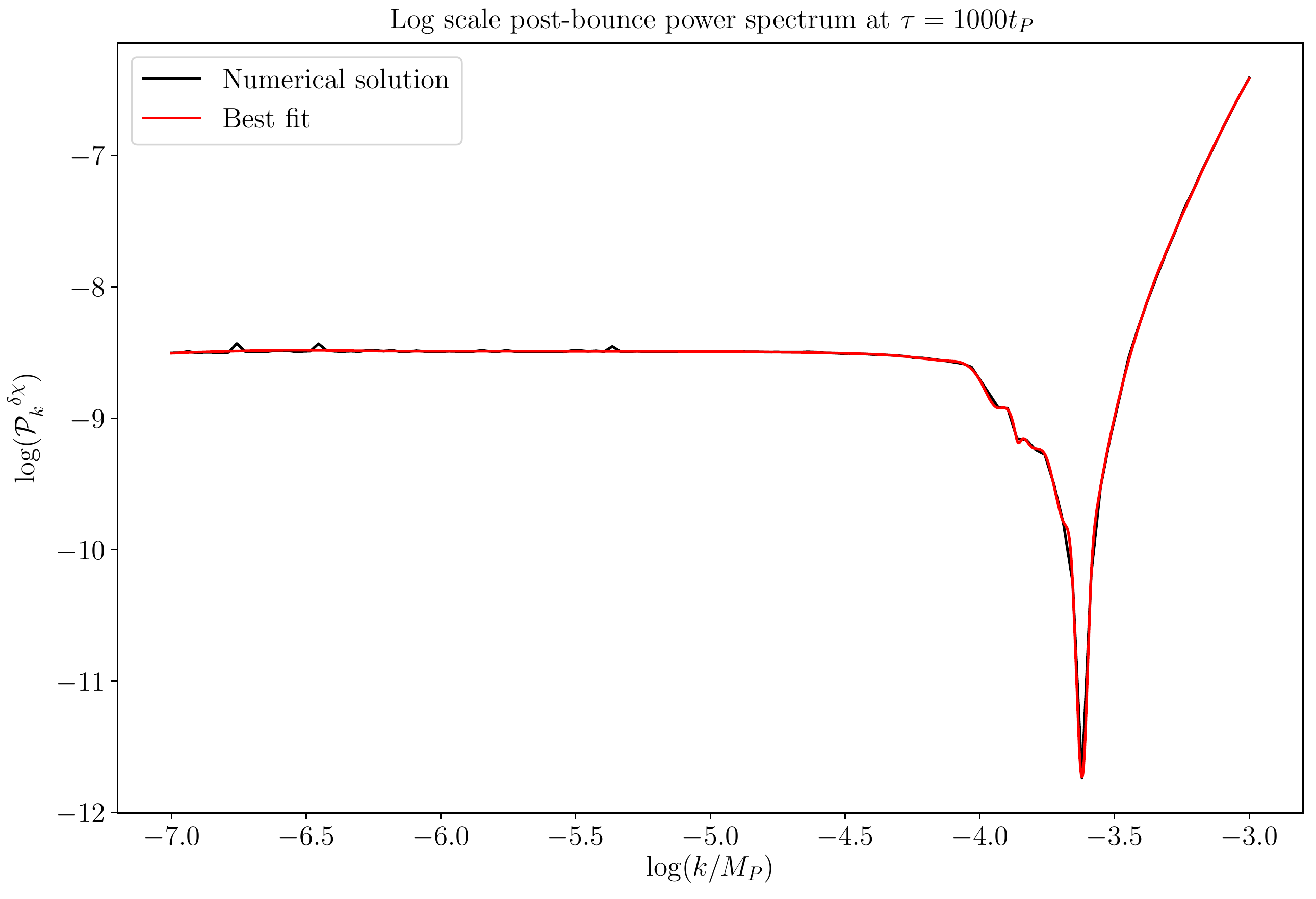}
\caption{\label{fig:post_bounce_ps_final} Post-bounce power spectrum for $\chi$-field perturbations evaluated at $\tau_f = 1000 t_P$, well after Cuscuton modifications become negligible again. The vacuum initial conditions are imposed at $\tau_i = -10^{-8} t_P$ and $\Lambda = 10^{-3.5}M_p$. The smooth red line represents a best fit through the numerical solution shown in black.}
\end{figure}
Figure \eqref{fig:post_bounce_ps_final} shows the power spectrum for the same entropy modes after they have gone through the bounce and evaluated at the time $\tau_f = 1000 t_P$. The black line is the actual numerical result and red curve shows the best fitted smooth function through the numerical result. As shown in the figure, the shape of the resulting power spectrum remains scale-invariant for $k\lesssim 10^{-4} M_{P}$ and is not impacted by the transition trough the bounce. In fact, if we assume the universe transitioned into radiation at this time, the actual modes of cosmological interest corresponding to CMB scales today would  be more than 10 orders of magnitudes smaller in $k$ than the range showed in this plot and are therefore even less impacted. Note that at this stage, from the model building perspective, our goal was simply to check if this model can produce a close to scale-invariant spectrum \footnote{We reiterate that in analogy to inflationary models, this is similar to first realizing that spectator scalar fields in de Sitter backgrounds generate scale invariant power-spectra. Then utilizing that observation and implementing a slowly rolling potential, inflationary models can accommodate quasi-de Sitter backgrounds where small deviations of $n_s$ from one is achieved by adjusting the derivatives of the potential.}. Testing for the proximity of the numerical solution to exact scale-invariance, in the range of $10^{-7} M_P \lesssim k\lesssim 10^{-4.5} M_P $, we obtain  $n_s - 1 \approx -0.0036$. Although this result should not be compared to the actual observational tilt\footnote{Current observation from PLANCK data \cite{Aghanim:2018eyx} require $n_s=0.965\pm 0.004$.}, it achieves our main goal of producing a solution which is nearly scale-invariant. As we mentioned before, when producing the actual observational tilt, one also needs to take into account additional effects such as the details of how the entropy modes are converted into adiabatic modes. The conversion itself requires some additional mechanism, such as premodulated preheating or the curvaton process. In Ekpyrotic models, the conversion of the entropy perturbations have been studied in detail and can happen either before the bounce, during the contracting phase \cite{Lehners:2007ac, Lehners:2009qu}, or after the bounce \cite{Fertig:2016czu}. We leave the study of these effects in Cuscuton bounce for future work.

As a final note to end this section, Figure \eqref{fig:post_bounce_ps_final} demonstrates the impact of the bounce transition on modes in the range $ 10^{-4} M_P \lesssim k \lesssim 10^{-3}M_P$ as well. Our result shows that for these modes, the deviation of $\frac{q''}{q}$ from $\frac{2}{\tau^2}$ in vicinity of $k^2\sim \frac{q''}{q}$ crossing is significant. However, as we pointed out earlier since these scales are near the Planck scale, in a non-inflationary universe they would not be observable on cosmological scales.

\section{Power spectrum of primordial gravitational waves in Cuscuton bounce}\label{PStensor}

\begin{figure}[tbp]
\centering 
\includegraphics[width=0.9\textwidth]{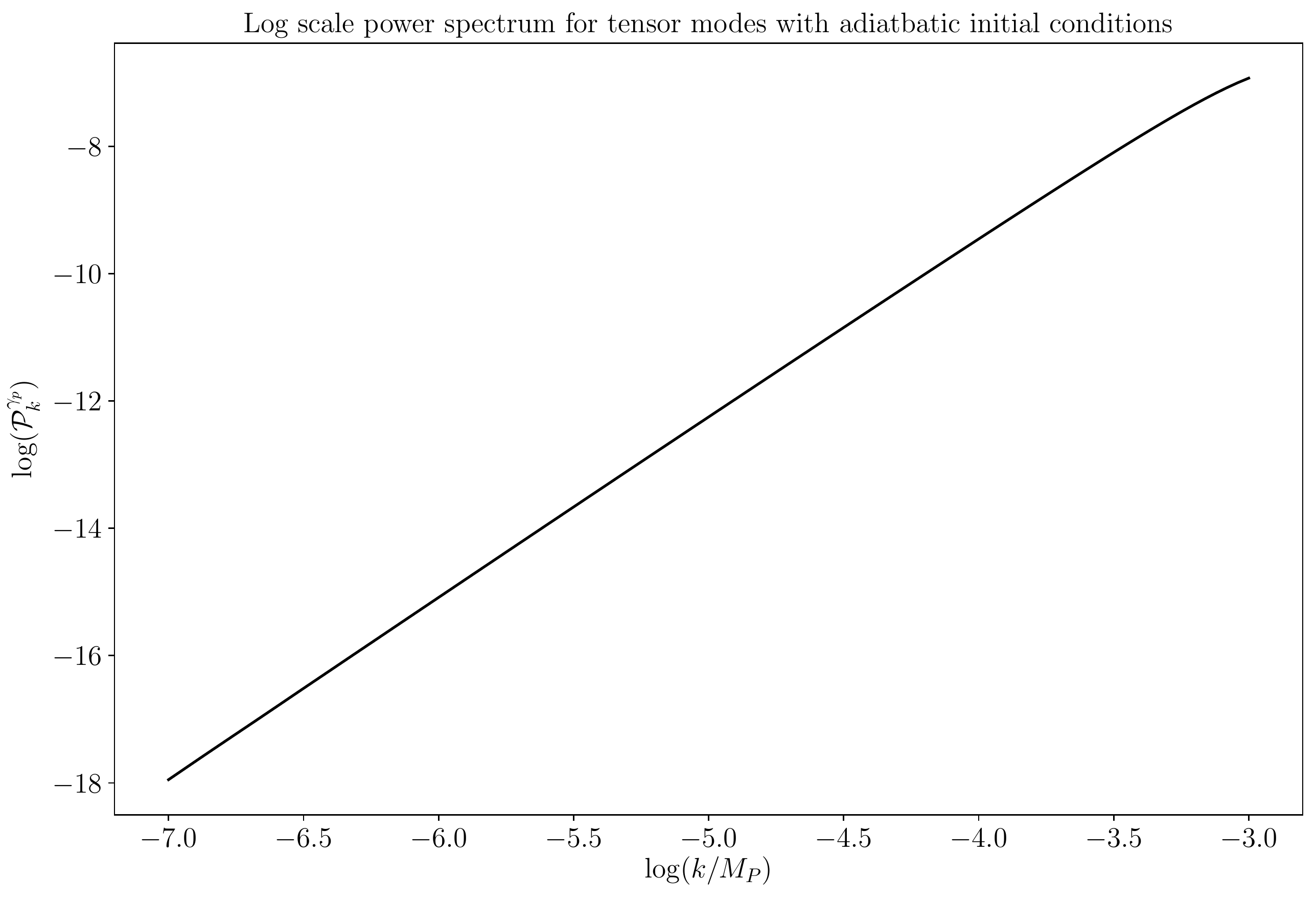}
\caption{\label{fig:ps_tensor_ad} Logarithmic scale power-spectrum for tensor perturbations with vacuum initial conditions set at $\tau_i = -10^{8} t_P$. The final time is taken as $\tau_f = 1000 t_P$. Power spectrum has the blue tilt of $n_t\sim 2$ with an amplitude which is many orders of magnitudes below observable thresholds at cosmological scales.}
\end{figure}
We now return back to tensor modes and look into the power spectrum of primordial gravitational waves generated in Cuscuton bounce. While primordial gravitational waves have not been detected yet, if ever detected, they could play a crucial role in distinguishing between inflationary and bouncing models. As has been argued in \cite{Geshnizjani:2014bya}, there are some basic arguments on why extracting enough power from the tensor quantum fluctuations at extremely small scales and converting them to observable tensor modes at cosmological scales relies on gaining a large number of e-foldings, which is possible during the inflationary epoch but by construction is absent in bounce scenarios. Therefore, generally one does not expect bounce models to produce observable primordial tensor modes. As we show here, the same argument applies to Cuscuton bounce. 

First, note that since $\chi$-field does not couple to metric perturbations at second order, the action for tensor modes given by \eqref{eq:gw_action2}, remains unchanged. Furthermore, since the background evolution for the single field and two field Cuscuton bounce are the same, the evolution of tensor modes in both scenarios are also identical. Once again, the quantization procedure can be carried out by introducing the canonical variable $v_p$ for both polarizations of the tensor modes \cite{Mukhanov:1990me, Brandenberger:2003vk},
\begin{align}
    v_p = \frac{a M_P }{ \sqrt{2}} \gamma_p, 
\end{align} 
so that the action in Fourier space can be rewritten as
\begin{align}
    S_{v_p}^{(2)} = \frac{1}{2} \sum_{p = \times, +} \int d\tau d^3k \ \left[ v_{p}'^2(k,\tau) - \left( k^2 - \frac{a''}{a} \right) v_{p}^2(k,\tau) \right],
\end{align}
and the equation of motion for each polarization $v_p$ is given by
\begin{align}\label{eomgw}
    v_p'' + \left( k^2 - \frac{a''}{a}  \right) v_p = 0.
\end{align}
Next, considering the adiabatic vacuum initial conditions by imposing 
\begin{align}
    v_p(k, \tau_i) = \frac{1}{\sqrt{2 k}} e^{-i k \tau_i} \qquad
    v_p'(k, \tau_i) = -i \sqrt{\frac{k}{2}} e^{-i k \tau_i},
\end{align}
at $\tau_i$ such that $k^2 \gg \frac{a''}{a}|_{\tau_i}$, we numerically solve equation \eqref{eomgw} and evaluate the power spectrum at some later time $\tau_f$, 
\begin{align}
    \mathcal{P}_k^{\gamma_p} (k, \tau_f) = \frac{k^3}{2 \pi^2} | \gamma_p (k, \tau_f) |^2 =  \frac{k^3}{2\pi^2} \frac{|2 v_p(k, \tau_f) | ^2}{ M_P^2 ~a^2(\tau_f) }.
\end{align}
Our result shows that tensor modes never freeze out in pre-bounce transition and they are not affected through the bounce either. Figure \eqref{fig:ps_tensor_ad} displays the logarithmic scale power spectrum for tensor modes evaluated post bounce, at $\tau_f = 1000 t_P$ and with vacuum initial conditions set at $\tau_i = -10^{8}t_p$. As expected, the produced power spectrum retains its vacuum sub-freezing tilt, i.e. $n_t\sim 2$ with an amplitude which is many orders of magnitudes below observable thresholds at cosmological scales.

\section{Conclusion}\label{conclusion}

In this work, we studied the power spectra of curvature, iso-curvature and tensor perturbations in the Cuscuton bounce proposed in \cite{Boruah:2018pvq}. For curvature perturbations, we considered three different initial conditions: standard adiabatic vacuum initial conditions, instantaneous minimal energy condition at bounce, as well as initial thermal condition. In all three cases, the generated power-spectrum for scalar perturbations were found to be strongly blue. Next, we investigated the possibility of a two field model, generating a nearly scale-invariant iso-curvature/entropy power spectrum for scalar perturbations. We found that this can be done with a very simple model with a spectator field that is kinetically coupled to the primary matter field.
This is a very interesting result since our solution does not rely on any {\it ad hoc} matching condition across the bounce or encounter any instability or singularities as it passes through the bounce. While this is a significant result, we note that to explain the observed tilt and amplitude of scalar perturbations \cite{Aghanim:2018eyx}, the model requires an additional phase that converts entropy modes into adiabatic modes, and similar to inflation, it requires a reheating phase for the universe to transition into the radiation phase. 
We leave the exploration of all these interesting aspects as well other potentially important features, such as non-Gaussianities or pre-bounce smoothing, for future work.

To complete our work on the generation of the power spectrum, we also investigated the behaviour of tensor modes in Cuscuton bounce. We first showed that Cuscuton bounce is stable under tensor perturbations as well. Then we obtained the power spectrum for primordial gravitational waves produced in a Cuscuton scenario and found that similar to other bounce models, the tensor index $n_t$ is strongly blue, and so the model does not predict an observable spectrum of primordial gravitational waves.

\acknowledgments

We would like to thank Hyung J. Kim and Jos\'e T. G\'alvez Ghersi for very useful discussions in the early stages of the project. We would also like to thank Erik Schnetter for his insight in the computational aspects of this project. In addition, we thank Niayesh Afshordi for reading through the manuscript on a short notice and providing his feedback. This project was supported by the Discovery Grant from the Natural Science and Engineering Research Council of Canada (NSERC). G.G. is supported partly by the Perimeter Institute (PI) as well. Research at PI is supported by the Government of Canada through the Department of Innovation, Science and Economic Development Canada and led by the Province of Ontario through the Ministry of Research, Innovation and Science.


\bibliographystyle{JHEP}
\bibliography{main.bib}

\providecommand{\href}[2]{#2}\begingroup\raggedright\begin{thebibliography}{100}

\bibitem{Guth:1980zm}
A.~H. Guth, \emph{{The Inflationary Universe: A Possible Solution to the
  Horizon and Flatness Problems}},
  \href{https://doi.org/10.1103/PhysRevD.23.347}{\emph{Adv. Ser. Astrophys.
  Cosmol.} {\bfseries 3} (1987) 139}.

\bibitem{Linde:1981mu}
A.~D. Linde, \emph{{A New Inflationary Universe Scenario: A Possible Solution
  of the Horizon, Flatness, Homogeneity, Isotropy and Primordial Monopole
  Problems}}, \href{https://doi.org/10.1016/0370-2693(82)91219-9}{\emph{Adv.
  Ser. Astrophys. Cosmol.} {\bfseries 3} (1987) 149}.

\bibitem{Senatore:2016aui}
L.~Senatore, \emph{{Lectures on Inflation}},  in \emph{{Theoretical Advanced
  Study Institute in Elementary Particle Physics}: {New Frontiers in Fields and
  Strings}}, pp.~447--543, 2017,
  \href{https://doi.org/10.1142/9789813149441_0008}{DOI}
  [\href{https://arxiv.org/abs/1609.00716}{{\ttfamily 1609.00716}}].

\bibitem{Hinshaw:2003ex}
{\scshape WMAP} collaboration, \emph{{First year Wilkinson Microwave Anisotropy
  Probe (WMAP) observations: The Angular power spectrum}},
  \href{https://doi.org/10.1086/377225}{\emph{Astrophys. J. Suppl.} {\bfseries
  148} (2003) 135} [\href{https://arxiv.org/abs/astro-ph/0302217}{{\ttfamily
  astro-ph/0302217}}].

\bibitem{Aghanim:2018eyx}
{\scshape Planck} collaboration, \emph{{Planck 2018 results. VI. Cosmological
  parameters}},
  \href{https://doi.org/10.1051/0004-6361/201833910}{\emph{Astron. Astrophys.}
  {\bfseries 641} (2020) A6}
  [\href{https://arxiv.org/abs/1807.06209}{{\ttfamily 1807.06209}}].

\bibitem{Alam:2020sor}
{\scshape eBOSS} collaboration, \emph{{The Completed SDSS-IV extended Baryon
  Oscillation Spectroscopic Survey: Cosmological Implications from two Decades
  of Spectroscopic Surveys at the Apache Point observatory}},
  \href{https://arxiv.org/abs/2007.08991}{{\ttfamily 2007.08991}}.

\bibitem{Akrami:2018odb}
{\scshape Planck} collaboration, \emph{{Planck 2018 results. X. Constraints on
  inflation}},  \href{https://arxiv.org/abs/1807.06211}{{\ttfamily
  1807.06211}}.

\bibitem{Babic:2019ify}
I.~Babic, C.~Burgess and G.~Geshnizjani, \emph{{Keeping an Eye on DBI:
  Power-counting for small-$c_s$ Cosmology}},
  \href{https://doi.org/10.1088/1475-7516/2020/05/023}{\emph{JCAP} {\bfseries
  05} (2020) 023} [\href{https://arxiv.org/abs/1910.05277}{{\ttfamily
  1910.05277}}].

\bibitem{Burgess:2017ytm}
C.~Burgess, \emph{{Intro to Effective Field Theories and Inflation}},
  \href{https://arxiv.org/abs/1711.10592}{{\ttfamily 1711.10592}}.

\bibitem{Martin:2000xs}
J.~Martin and R.~H. Brandenberger, \emph{{The TransPlanckian problem of
  inflationary cosmology}},
  \href{https://doi.org/10.1103/PhysRevD.63.123501}{\emph{Phys. Rev. D}
  {\bfseries 63} (2001) 123501}
  [\href{https://arxiv.org/abs/hep-th/0005209}{{\ttfamily hep-th/0005209}}].

\bibitem{Brandenberger:2000wr}
R.~H. Brandenberger and J.~Martin, \emph{{The Robustness of inflation to
  changes in superPlanck scale physics}},
  \href{https://doi.org/10.1142/S0217732301004170}{\emph{Mod. Phys. Lett. A}
  {\bfseries 16} (2001) 999}
  [\href{https://arxiv.org/abs/astro-ph/0005432}{{\ttfamily
  astro-ph/0005432}}].

\bibitem{Brandenberger:2012aj}
R.~H. Brandenberger and J.~Martin, \emph{{Trans-Planckian Issues for
  Inflationary Cosmology}},
  \href{https://doi.org/10.1088/0264-9381/30/11/113001}{\emph{Class. Quant.
  Grav.} {\bfseries 30} (2013) 113001}
  [\href{https://arxiv.org/abs/1211.6753}{{\ttfamily 1211.6753}}].

\bibitem{Ashoorioon:2018uey}
A.~Ashoorioon, R.~Casadio, M.~Cicoli, G.~Geshnizjani and H.~J. Kim,
  \emph{{Extended Effective Field Theory of Inflation}},
  \href{https://doi.org/10.1007/JHEP02(2018)172}{\emph{JHEP} {\bfseries 02}
  (2018) 172} [\href{https://arxiv.org/abs/1802.03040}{{\ttfamily
  1802.03040}}].

\bibitem{Bedroya:2019snp}
A.~Bedroya and C.~Vafa, \emph{{Trans-Planckian Censorship and the Swampland}},
  \href{https://arxiv.org/abs/1909.11063}{{\ttfamily 1909.11063}}.

\bibitem{Bedroya:2019tba}
A.~Bedroya, R.~Brandenberger, M.~Loverde and C.~Vafa, \emph{{Trans-Planckian
  Censorship and Inflationary Cosmology}},
  \href{https://doi.org/10.1103/PhysRevD.101.103502}{\emph{Phys. Rev. D}
  {\bfseries 101} (2020) 103502}
  [\href{https://arxiv.org/abs/1909.11106}{{\ttfamily 1909.11106}}].

\bibitem{Ooguri:2006in}
H.~Ooguri and C.~Vafa, \emph{{On the Geometry of the String Landscape and the
  Swampland}},
  \href{https://doi.org/10.1016/j.nuclphysb.2006.10.033}{\emph{Nucl. Phys. B}
  {\bfseries 766} (2007) 21}
  [\href{https://arxiv.org/abs/hep-th/0605264}{{\ttfamily hep-th/0605264}}].

\bibitem{Brennan:2017rbf}
T.~D. Brennan, F.~Carta and C.~Vafa, \emph{{The String Landscape, the
  Swampland, and the Missing Corner}},
  \href{https://doi.org/10.22323/1.305.0015}{\emph{PoS} {\bfseries TASI2017}
  (2017) 015} [\href{https://arxiv.org/abs/1711.00864}{{\ttfamily
  1711.00864}}].

\bibitem{Obied:2018sgi}
G.~Obied, H.~Ooguri, L.~Spodyneiko and C.~Vafa, \emph{{De Sitter Space and the
  Swampland}},  \href{https://arxiv.org/abs/1806.08362}{{\ttfamily
  1806.08362}}.

\bibitem{Palti:2019pca}
E.~Palti, \emph{{The Swampland: Introduction and Review}},
  \href{https://doi.org/10.1002/prop.201900037}{\emph{Fortsch. Phys.}
  {\bfseries 67} (2019) 1900037}
  [\href{https://arxiv.org/abs/1903.06239}{{\ttfamily 1903.06239}}].

\bibitem{Achucarro:2018vey}
A.~Achúcarro and G.~A. Palma, \emph{{The string swampland constraints require
  multi-field inflation}},
  \href{https://doi.org/10.1088/1475-7516/2019/02/041}{\emph{JCAP} {\bfseries
  02} (2019) 041} [\href{https://arxiv.org/abs/1807.04390}{{\ttfamily
  1807.04390}}].

\bibitem{Dias:2018ngv}
M.~Dias, J.~Frazer, A.~Retolaza and A.~Westphal, \emph{{Primordial
  Gravitational Waves and the Swampland}},
  \href{https://doi.org/10.1002/prop.201800063}{\emph{Fortsch. Phys.}
  {\bfseries 67} (2019) 2} [\href{https://arxiv.org/abs/1807.06579}{{\ttfamily
  1807.06579}}].

\bibitem{Agrawal:2018own}
P.~Agrawal, G.~Obied, P.~J. Steinhardt and C.~Vafa, \emph{{On the Cosmological
  Implications of the String Swampland}},
  \href{https://doi.org/10.1016/j.physletb.2018.07.040}{\emph{Phys. Lett. B}
  {\bfseries 784} (2018) 271}
  [\href{https://arxiv.org/abs/1806.09718}{{\ttfamily 1806.09718}}].

\bibitem{Kinney:2018nny}
W.~H. Kinney, S.~Vagnozzi and L.~Visinelli, \emph{{The zoo plot meets the
  swampland: mutual (in)consistency of single-field inflation, string
  conjectures, and cosmological data}},
  \href{https://doi.org/10.1088/1361-6382/ab1d87}{\emph{Class. Quant. Grav.}
  {\bfseries 36} (2019) 117001}
  [\href{https://arxiv.org/abs/1808.06424}{{\ttfamily 1808.06424}}].

\bibitem{Lin:2019pmj}
W.-C. Lin and W.~H. Kinney, \emph{{Trans-Planckian Censorship and
  $k$-inflation}},
  \href{https://doi.org/10.1103/PhysRevD.101.123534}{\emph{Phys. Rev. D}
  {\bfseries 101} (2020) 123534}
  [\href{https://arxiv.org/abs/1911.03736}{{\ttfamily 1911.03736}}].

\bibitem{Garfinkle:2008ei}
D.~Garfinkle, W.~C. Lim, F.~Pretorius and P.~J. Steinhardt, \emph{{Evolution to
  a smooth universe in an ekpyrotic contracting phase with w > 1}},
  \href{https://doi.org/10.1103/PhysRevD.78.083537}{\emph{Phys. Rev. D}
  {\bfseries 78} (2008) 083537}
  [\href{https://arxiv.org/abs/0808.0542}{{\ttfamily 0808.0542}}].

\bibitem{Cook:2020oaj}
W.~G. Cook, I.~A. Glushchenko, A.~Ijjas, F.~Pretorius and P.~J. Steinhardt,
  \emph{{Supersmoothing through Slow Contraction}},
  \href{https://arxiv.org/abs/2006.01172}{{\ttfamily 2006.01172}}.

\bibitem{Ijjas:2020dws}
A.~Ijjas, W.~G. Cook, F.~Pretorius, P.~J. Steinhardt and E.~Y. Davies,
  \emph{{Robustness of slow contraction to cosmic initial conditions}},
  \href{https://arxiv.org/abs/2006.04999}{{\ttfamily 2006.04999}}.

\bibitem{steinhardt1982natural}
P.~J. Steinhardt, \emph{Natural inflation},  in \emph{The Very Early Universe,
  ed. by G. Gibbons, S. Hawking and S. Siklos}, p.~251–66, Cambridge
  University Press, (1983).

\bibitem{Vilenkin:1983xq}
A.~Vilenkin, \emph{{The Birth of Inflationary Universes}},
  \href{https://doi.org/10.1103/PhysRevD.27.2848}{\emph{Phys. Rev. D}
  {\bfseries 27} (1983) 2848}.

\bibitem{Linde:1986fc}
A.~D. Linde, \emph{{Eternal Chaotic Inflation}},
  \href{https://doi.org/10.1142/S0217732386000129}{\emph{Mod. Phys. Lett. A}
  {\bfseries 1} (1986) 81}.

\bibitem{Guth:2007ng}
A.~H. Guth, \emph{{Eternal inflation and its implications}},
  \href{https://doi.org/10.1088/1751-8113/40/25/S25}{\emph{J. Phys. A}
  {\bfseries 40} (2007) 6811}
  [\href{https://arxiv.org/abs/hep-th/0702178}{{\ttfamily hep-th/0702178}}].

\bibitem{Borde:1993xh}
A.~Borde and A.~Vilenkin, \emph{{Eternal inflation and the initial
  singularity}}, \href{https://doi.org/10.1103/PhysRevLett.72.3305}{\emph{Phys.
  Rev. Lett.} {\bfseries 72} (1994) 3305}
  [\href{https://arxiv.org/abs/gr-qc/9312022}{{\ttfamily gr-qc/9312022}}].

\bibitem{Geshnizjani:2011dk}
G.~Geshnizjani, W.~H. Kinney and A.~Moradinezhad~Dizgah, \emph{{General
  conditions for scale-invariant perturbations in an expanding universe}},
  \href{https://doi.org/10.1088/1475-7516/2011/11/049}{\emph{JCAP} {\bfseries
  11} (2011) 049} [\href{https://arxiv.org/abs/1107.1241}{{\ttfamily
  1107.1241}}].

\bibitem{Geshnizjani:2013lza}
G.~Geshnizjani and N.~Ahmadi, \emph{{Can non-local or higher derivative
  theories provide alternatives to inflation?}},
  \href{https://doi.org/10.1088/1475-7516/2013/11/029}{\emph{JCAP} {\bfseries
  11} (2013) 029} [\href{https://arxiv.org/abs/1309.4782}{{\ttfamily
  1309.4782}}].

\bibitem{Geshnizjani:2014bya}
G.~Geshnizjani and W.~H. Kinney, \emph{{Theoretical implications of detecting
  gravitational waves}},
  \href{https://doi.org/10.1088/1475-7516/2015/08/008}{\emph{JCAP} {\bfseries
  08} (2015) 008} [\href{https://arxiv.org/abs/1410.4968}{{\ttfamily
  1410.4968}}].

\bibitem{Dvali:2020cgt}
G.~Dvali, A.~Kehagias and A.~Riotto, \emph{{Inflation and Decoupling}},
  \href{https://arxiv.org/abs/2005.05146}{{\ttfamily 2005.05146}}.

\bibitem{Finelli:2001sr}
F.~Finelli and R.~Brandenberger, \emph{{On the generation of a scale invariant
  spectrum of adiabatic fluctuations in cosmological models with a contracting
  phase}}, \href{https://doi.org/10.1103/PhysRevD.65.103522}{\emph{Phys. Rev.
  D} {\bfseries 65} (2002) 103522}
  [\href{https://arxiv.org/abs/hep-th/0112249}{{\ttfamily hep-th/0112249}}].

\bibitem{Brandenberger:1988aj}
R.~H. Brandenberger and C.~Vafa, \emph{{Superstrings in the Early Universe}},
  \href{https://doi.org/10.1016/0550-3213(89)90037-0}{\emph{Nucl. Phys. B}
  {\bfseries 316} (1989) 391}.

\bibitem{Qiu:2011cy}
T.~Qiu, J.~Evslin, Y.-F. Cai, M.~Li and X.~Zhang, \emph{{Bouncing Galileon
  Cosmologies}},
  \href{https://doi.org/10.1088/1475-7516/2011/10/036}{\emph{JCAP} {\bfseries
  10} (2011) 036} [\href{https://arxiv.org/abs/1108.0593}{{\ttfamily
  1108.0593}}].

\bibitem{Cai:2007zv}
Y.-F. Cai, T.~Qiu, R.~Brandenberger, Y.-S. Piao and X.~Zhang, \emph{{On
  Perturbations of Quintom Bounce}},
  \href{https://doi.org/10.1088/1475-7516/2008/03/013}{\emph{JCAP} {\bfseries
  03} (2008) 013} [\href{https://arxiv.org/abs/0711.2187}{{\ttfamily
  0711.2187}}].

\bibitem{Cai:2012va}
Y.-F. Cai, D.~A. Easson and R.~Brandenberger, \emph{{Towards a Nonsingular
  Bouncing Cosmology}},
  \href{https://doi.org/10.1088/1475-7516/2012/08/020}{\emph{JCAP} {\bfseries
  08} (2012) 020} [\href{https://arxiv.org/abs/1206.2382}{{\ttfamily
  1206.2382}}].

\bibitem{Cai:2013kja}
Y.-F. Cai, E.~McDonough, F.~Duplessis and R.~H. Brandenberger, \emph{{Two Field
  Matter Bounce Cosmology}},
  \href{https://doi.org/10.1088/1475-7516/2013/10/024}{\emph{JCAP} {\bfseries
  10} (2013) 024} [\href{https://arxiv.org/abs/1305.5259}{{\ttfamily
  1305.5259}}].

\bibitem{Gasperini:1992em}
M.~Gasperini and G.~Veneziano, \emph{{Pre - big bang in string cosmology}},
  \href{https://doi.org/10.1016/0927-6505(93)90017-8}{\emph{Astropart. Phys.}
  {\bfseries 1} (1993) 317}
  [\href{https://arxiv.org/abs/hep-th/9211021}{{\ttfamily hep-th/9211021}}].

\bibitem{Cai:2016thi}
Y.~Cai, Y.~Wan, H.-G. Li, T.~Qiu and Y.-S. Piao, \emph{{The Effective Field
  Theory of nonsingular cosmology}},
  \href{https://doi.org/10.1007/JHEP01(2017)090}{\emph{JHEP} {\bfseries 01}
  (2017) 090} [\href{https://arxiv.org/abs/1610.03400}{{\ttfamily
  1610.03400}}].

\bibitem{Cai:2017tku}
Y.~Cai, H.-G. Li, T.~Qiu and Y.-S. Piao, \emph{{The Effective Field Theory of
  nonsingular cosmology: II}},
  \href{https://doi.org/10.1140/epjc/s10052-017-4938-y}{\emph{Eur. Phys. J. C}
  {\bfseries 77} (2017) 369}
  [\href{https://arxiv.org/abs/1701.04330}{{\ttfamily 1701.04330}}].

\bibitem{Cai:2017dyi}
Y.~Cai and Y.-S. Piao, \emph{{A covariant Lagrangian for stable nonsingular
  bounce}}, \href{https://doi.org/10.1007/JHEP09(2017)027}{\emph{JHEP}
  {\bfseries 09} (2017) 027}
  [\href{https://arxiv.org/abs/1705.03401}{{\ttfamily 1705.03401}}].

\bibitem{Lehners:2008vx}
J.-L. Lehners, \emph{{Ekpyrotic and Cyclic Cosmology}},
  \href{https://doi.org/10.1016/j.physrep.2008.06.001}{\emph{Phys. Rept.}
  {\bfseries 465} (2008) 223}
  [\href{https://arxiv.org/abs/0806.1245}{{\ttfamily 0806.1245}}].

\bibitem{Khoury:2001wf}
J.~Khoury, B.~A. Ovrut, P.~J. Steinhardt and N.~Turok, \emph{{The Ekpyrotic
  universe: Colliding branes and the origin of the hot big bang}},
  \href{https://doi.org/10.1103/PhysRevD.64.123522}{\emph{Phys. Rev. D}
  {\bfseries 64} (2001) 123522}
  [\href{https://arxiv.org/abs/hep-th/0103239}{{\ttfamily hep-th/0103239}}].

\bibitem{Ijjas:2016vtq}
A.~Ijjas and P.~J. Steinhardt, \emph{{Fully stable cosmological solutions with
  a non-singular classical bounce}},
  \href{https://doi.org/10.1016/j.physletb.2016.11.047}{\emph{Phys. Lett. B}
  {\bfseries 764} (2017) 289}
  [\href{https://arxiv.org/abs/1609.01253}{{\ttfamily 1609.01253}}].

\bibitem{Ijjas:2016tpn}
A.~Ijjas and P.~J. Steinhardt, \emph{{Classically stable nonsingular
  cosmological bounces}},
  \href{https://doi.org/10.1103/PhysRevLett.117.121304}{\emph{Phys. Rev. Lett.}
  {\bfseries 117} (2016) 121304}
  [\href{https://arxiv.org/abs/1606.08880}{{\ttfamily 1606.08880}}].

\bibitem{Dobre:2017pnt}
D.~A. Dobre, A.~V. Frolov, J.~T. G\'alvez~Ghersi, S.~Ramazanov and A.~Vikman,
  \emph{{Unbraiding the Bounce: Superluminality around the Corner}},
  \href{https://doi.org/10.1088/1475-7516/2018/03/020}{\emph{JCAP} {\bfseries
  03} (2018) 020} [\href{https://arxiv.org/abs/1712.10272}{{\ttfamily
  1712.10272}}].

\bibitem{Easson:2011zy}
D.~A. Easson, I.~Sawicki and A.~Vikman, \emph{{G-Bounce}},
  \href{https://doi.org/10.1088/1475-7516/2011/11/021}{\emph{JCAP} {\bfseries
  11} (2011) 021} [\href{https://arxiv.org/abs/1109.1047}{{\ttfamily
  1109.1047}}].

\bibitem{Creminelli:2006xe}
P.~Creminelli, M.~A. Luty, A.~Nicolis and L.~Senatore, \emph{{Starting the
  Universe: Stable Violation of the Null Energy Condition and Non-standard
  Cosmologies}},
  \href{https://doi.org/10.1088/1126-6708/2006/12/080}{\emph{JHEP} {\bfseries
  12} (2006) 080} [\href{https://arxiv.org/abs/hep-th/0606090}{{\ttfamily
  hep-th/0606090}}].

\bibitem{Creminelli:2007aq}
P.~Creminelli and L.~Senatore, \emph{{A Smooth bouncing cosmology with scale
  invariant spectrum}},
  \href{https://doi.org/10.1088/1475-7516/2007/11/010}{\emph{JCAP} {\bfseries
  11} (2007) 010} [\href{https://arxiv.org/abs/hep-th/0702165}{{\ttfamily
  hep-th/0702165}}].

\bibitem{Creminelli:2016zwa}
P.~Creminelli, D.~Pirtskhalava, L.~Santoni and E.~Trincherini, \emph{{Stability
  of Geodesically Complete Cosmologies}},
  \href{https://doi.org/10.1088/1475-7516/2016/11/047}{\emph{JCAP} {\bfseries
  11} (2016) 047} [\href{https://arxiv.org/abs/1610.04207}{{\ttfamily
  1610.04207}}].

\bibitem{Dubovsky:2005xd}
S.~Dubovsky, T.~Gregoire, A.~Nicolis and R.~Rattazzi, \emph{{Null energy
  condition and superluminal propagation}},
  \href{https://doi.org/10.1088/1126-6708/2006/03/025}{\emph{JHEP} {\bfseries
  03} (2006) 025} [\href{https://arxiv.org/abs/hep-th/0512260}{{\ttfamily
  hep-th/0512260}}].

\bibitem{Sawicki:2012pz}
I.~Sawicki and A.~Vikman, \emph{{Hidden Negative Energies in Strongly
  Accelerated Universes}},
  \href{https://doi.org/10.1103/PhysRevD.87.067301}{\emph{Phys. Rev. D}
  {\bfseries 87} (2013) 067301}
  [\href{https://arxiv.org/abs/1209.2961}{{\ttfamily 1209.2961}}].

\bibitem{Rubakov:2014jja}
V.~Rubakov, \emph{{The Null Energy Condition and its violation}},
  \href{https://doi.org/10.3367/UFNe.0184.201402b.0137}{\emph{Usp. Fiz. Nauk}
  {\bfseries 184} (2014) 137}
  [\href{https://arxiv.org/abs/1401.4024}{{\ttfamily 1401.4024}}].

\bibitem{Libanov:2016kfc}
M.~Libanov, S.~Mironov and V.~Rubakov, \emph{{Generalized Galileons:
  instabilities of bouncing and Genesis cosmologies and modified Genesis}},
  \href{https://doi.org/10.1088/1475-7516/2016/08/037}{\emph{JCAP} {\bfseries
  08} (2016) 037} [\href{https://arxiv.org/abs/1605.05992}{{\ttfamily
  1605.05992}}].

\bibitem{Kobayashi:2016xpl}
T.~Kobayashi, \emph{{Generic instabilities of nonsingular cosmologies in
  Horndeski theory: A no-go theorem}},
  \href{https://doi.org/10.1103/PhysRevD.94.043511}{\emph{Phys. Rev. D}
  {\bfseries 94} (2016) 043511}
  [\href{https://arxiv.org/abs/1606.05831}{{\ttfamily 1606.05831}}].

\bibitem{Babichev:2007dw}
E.~Babichev, V.~Mukhanov and A.~Vikman, \emph{{k-Essence, superluminal
  propagation, causality and emergent geometry}},
  \href{https://doi.org/10.1088/1126-6708/2008/02/101}{\emph{JHEP} {\bfseries
  02} (2008) 101} [\href{https://arxiv.org/abs/0708.0561}{{\ttfamily
  0708.0561}}].

\bibitem{Adams:2006sv}
A.~Adams, N.~Arkani-Hamed, S.~Dubovsky, A.~Nicolis and R.~Rattazzi,
  \emph{{Causality, analyticity and an IR obstruction to UV completion}},
  \href{https://doi.org/10.1088/1126-6708/2006/10/014}{\emph{JHEP} {\bfseries
  10} (2006) 014} [\href{https://arxiv.org/abs/hep-th/0602178}{{\ttfamily
  hep-th/0602178}}].

\bibitem{Easson:2013bda}
D.~A. Easson, I.~Sawicki and A.~Vikman, \emph{{When Matter Matters}},
  \href{https://doi.org/10.1088/1475-7516/2013/07/014}{\emph{JCAP} {\bfseries
  07} (2013) 014} [\href{https://arxiv.org/abs/1304.3903}{{\ttfamily
  1304.3903}}].

\bibitem{Mironov:2019haz}
S.~Mironov, V.~Rubakov and V.~Volkova, \emph{{Cosmological scenarios with
  bounce and Genesis in Horndeski theory and beyond: An essay in honor of I.M.
  Khalatnikov on the occasion of his 100th birthday}},
  \href{https://arxiv.org/abs/1906.12139}{{\ttfamily 1906.12139}}.

\bibitem{Mironov:2019mye}
S.~Mironov, V.~Rubakov and V.~Volkova, \emph{{Subluminal cosmological bounce
  beyond Horndeski}},
  \href{https://doi.org/10.1088/1475-7516/2020/05/024}{\emph{JCAP} {\bfseries
  05} (2020) 024} [\href{https://arxiv.org/abs/1910.07019}{{\ttfamily
  1910.07019}}].

\bibitem{Mironov:2020pqh}
S.~Mironov, V.~Rubakov and V.~Volkova, \emph{{Superluminality in DHOST theory
  with extra scalar}},  \href{https://arxiv.org/abs/2011.14912}{{\ttfamily
  2011.14912}}.

\bibitem{Boruah:2017tvg}
S.~S. Boruah, H.~J. Kim and G.~Geshnizjani, \emph{{Theory of Cosmological
  Perturbations with Cuscuton}},
  \href{https://doi.org/10.1088/1475-7516/2017/07/022}{\emph{JCAP} {\bfseries
  07} (2017) 022} [\href{https://arxiv.org/abs/1704.01131}{{\ttfamily
  1704.01131}}].

\bibitem{Boruah:2018pvq}
S.~S. Boruah, H.~J. Kim, M.~Rouben and G.~Geshnizjani, \emph{{Cuscuton
  bounce}}, \href{https://doi.org/10.1088/1475-7516/2018/08/031}{\emph{JCAP}
  {\bfseries 08} (2018) 031}
  [\href{https://arxiv.org/abs/1802.06818}{{\ttfamily 1802.06818}}].

\bibitem{Afshordi:2006ad}
N.~Afshordi, D.~J. Chung and G.~Geshnizjani, \emph{{Cuscuton: A Causal Field
  Theory with an Infinite Speed of Sound}},
  \href{https://doi.org/10.1103/PhysRevD.75.083513}{\emph{Phys. Rev. D}
  {\bfseries 75} (2007) 083513}
  [\href{https://arxiv.org/abs/hep-th/0609150}{{\ttfamily hep-th/0609150}}].

\bibitem{Afshordi:2007yx}
N.~Afshordi, D.~J. Chung, M.~Doran and G.~Geshnizjani, \emph{{Cuscuton
  Cosmology: Dark Energy meets Modified Gravity}},
  \href{https://doi.org/10.1103/PhysRevD.75.123509}{\emph{Phys. Rev. D}
  {\bfseries 75} (2007) 123509}
  [\href{https://arxiv.org/abs/astro-ph/0702002}{{\ttfamily
  astro-ph/0702002}}].

\bibitem{Gomes:2017tzd}
H.~Gomes and D.~C. Guariento, \emph{{Hamiltonian analysis of the cuscuton}},
  \href{https://doi.org/10.1103/PhysRevD.95.104049}{\emph{Phys. Rev. D}
  {\bfseries 95} (2017) 104049}
  [\href{https://arxiv.org/abs/1703.08226}{{\ttfamily 1703.08226}}].

\bibitem{Quintin:2019orx}
J.~Quintin and D.~Yoshida, \emph{{Cuscuton gravity as a classically stable
  limiting curvature theory}},
  \href{https://doi.org/10.1088/1475-7516/2020/02/016}{\emph{JCAP} {\bfseries
  02} (2020) 016} [\href{https://arxiv.org/abs/1911.06040}{{\ttfamily
  1911.06040}}].

\bibitem{Iyonaga:2018vnu}
A.~Iyonaga, K.~Takahashi and T.~Kobayashi, \emph{{Extended Cuscuton:
  Formulation}},
  \href{https://doi.org/10.1088/1475-7516/2018/12/002}{\emph{JCAP} {\bfseries
  12} (2018) 002} [\href{https://arxiv.org/abs/1809.10935}{{\ttfamily
  1809.10935}}].

\bibitem{Belinsky:1970ew}
V.~Belinsky, I.~Khalatnikov and E.~Lifshitz, \emph{{Oscillatory approach to a
  singular point in the relativistic cosmology}},
  \href{https://doi.org/10.1080/00018737000101171}{\emph{Adv. Phys.} {\bfseries
  19} (1970) 525}.

\bibitem{Lifshitz:1963ps}
E.~Lifshitz and I.~Khalatnikov, \emph{{Investigations in relativistic
  cosmology}}, \href{https://doi.org/10.1080/00018736300101283}{\emph{Adv.
  Phys.} {\bfseries 12} (1963) 185}.

\bibitem{Battefeld:2014uga}
D.~Battefeld and P.~Peter, \emph{{A Critical Review of Classical Bouncing
  Cosmologies}},
  \href{https://doi.org/10.1016/j.physrep.2014.12.004}{\emph{Phys. Rept.}
  {\bfseries 571} (2015) 1} [\href{https://arxiv.org/abs/1406.2790}{{\ttfamily
  1406.2790}}].

\bibitem{Arnowitt:1962hi}
R.~L. Arnowitt, S.~Deser and C.~W. Misner, \emph{{The Dynamics of general
  relativity}}, \href{https://doi.org/10.1007/s10714-008-0661-1}{\emph{Gen.
  Rel. Grav.} {\bfseries 40} (2008) 1997}
  [\href{https://arxiv.org/abs/gr-qc/0405109}{{\ttfamily gr-qc/0405109}}].

\bibitem{Poisson:2009pwt}
E.~Poisson, \emph{{A Relativist's Toolkit: The Mathematics of Black-Hole
  Mechanics}}. Cambridge University Press, 12, 2009,
  \href{https://doi.org/10.1017/CBO9780511606601}{10.1017/CBO9780511606601}.

\bibitem{Mukhanov:2007zz}
V.~Mukhanov and S.~Winitzki, \emph{{Introduction to quantum effects in
  gravity}}. Cambridge University Press, 6, 2007.

\bibitem{Birrell:1982ix}
N.~Birrell and P.~Davies, \emph{{Quantum Fields in Curved Space}}, Cambridge
  Monographs on Mathematical Physics. Cambridge Univ. Press, Cambridge, UK, 2,
  1984,
  \href{https://doi.org/10.1017/CBO9780511622632}{10.1017/CBO9780511622632}.

\bibitem{Boyle:2018tzc}
L.~Boyle, K.~Finn and N.~Turok, \emph{{CPT-Symmetric Universe}},
  \href{https://doi.org/10.1103/PhysRevLett.121.251301}{\emph{Phys. Rev. Lett.}
  {\bfseries 121} (2018) 251301}
  [\href{https://arxiv.org/abs/1803.08928}{{\ttfamily 1803.08928}}].

\bibitem{Magueijo:2002pg}
J.~Magueijo and L.~Pogosian, \emph{{Could thermal fluctuations seed cosmic
  structure?}}, \href{https://doi.org/10.1103/PhysRevD.67.043518}{\emph{Phys.
  Rev. D} {\bfseries 67} (2003) 043518}
  [\href{https://arxiv.org/abs/astro-ph/0211337}{{\ttfamily
  astro-ph/0211337}}].

\bibitem{Ferreira:2007cb}
P.~Ferreira and J.~Magueijo, \emph{{Observing the temperature of the Big Bang
  through large scale structure}},
  \href{https://doi.org/10.1103/PhysRevD.78.061301}{\emph{Phys. Rev. D}
  {\bfseries 78} (2008) 061301}
  [\href{https://arxiv.org/abs/0708.0429}{{\ttfamily 0708.0429}}].

\bibitem{Magueijo:2008pm}
J.~Magueijo, \emph{{Speedy sound and cosmic structure}},
  \href{https://doi.org/10.1103/PhysRevLett.100.231302}{\emph{Phys. Rev. Lett.}
  {\bfseries 100} (2008) 231302}
  [\href{https://arxiv.org/abs/0803.0859}{{\ttfamily 0803.0859}}].

\bibitem{Agarwal:2014ona}
A.~Agarwal and N.~Afshordi, \emph{{Thermal Tachyacoustic Cosmology}},
  \href{https://doi.org/10.1103/PhysRevD.90.043528}{\emph{Phys. Rev. D}
  {\bfseries 90} (2014) 043528}
  [\href{https://arxiv.org/abs/1406.0575}{{\ttfamily 1406.0575}}].

\bibitem{Gasperini:2002bn}
M.~Gasperini and G.~Veneziano, \emph{{The Pre - big bang scenario in string
  cosmology}}, \href{https://doi.org/10.1016/S0370-1573(02)00389-7}{\emph{Phys.
  Rept.} {\bfseries 373} (2003) 1}
  [\href{https://arxiv.org/abs/hep-th/0207130}{{\ttfamily hep-th/0207130}}].

\bibitem{Creminelli:2004jg}
P.~Creminelli, A.~Nicolis and M.~Zaldarriaga, \emph{{Perturbations in bouncing
  cosmologies: Dynamical attractor versus scale invariance}},
  \href{https://doi.org/10.1103/PhysRevD.71.063505}{\emph{Phys. Rev. D}
  {\bfseries 71} (2005) 063505}
  [\href{https://arxiv.org/abs/hep-th/0411270}{{\ttfamily hep-th/0411270}}].

\bibitem{Tseng:2012qd}
C.-Y. Tseng, \emph{{Decoherence problem in an ekpyrotic phase}},
  \href{https://doi.org/10.1103/PhysRevD.87.023518}{\emph{Phys. Rev. D}
  {\bfseries 87} (2013) 023518}
  [\href{https://arxiv.org/abs/1210.0581}{{\ttfamily 1210.0581}}].

\bibitem{Battarra:2013cha}
L.~Battarra and J.-L. Lehners, \emph{{Quantum-to-classical transition for
  ekpyrotic perturbations}},
  \href{https://doi.org/10.1103/PhysRevD.89.063516}{\emph{Phys. Rev. D}
  {\bfseries 89} (2014) 063516}
  [\href{https://arxiv.org/abs/1309.2281}{{\ttfamily 1309.2281}}].

\bibitem{Buchbinder:2007ad}
E.~I. Buchbinder, J.~Khoury and B.~A. Ovrut, \emph{{New Ekpyrotic cosmology}},
  \href{https://doi.org/10.1103/PhysRevD.76.123503}{\emph{Phys. Rev. D}
  {\bfseries 76} (2007) 123503}
  [\href{https://arxiv.org/abs/hep-th/0702154}{{\ttfamily hep-th/0702154}}].

\bibitem{Notari:2002yc}
A.~Notari and A.~Riotto, \emph{{Isocurvature perturbations in the ekpyrotic
  universe}}, \href{https://doi.org/10.1016/S0550-3213(02)00765-4}{\emph{Nucl.
  Phys. B} {\bfseries 644} (2002) 371}
  [\href{https://arxiv.org/abs/hep-th/0205019}{{\ttfamily hep-th/0205019}}].

\bibitem{Lehners:2007ac}
J.-L. Lehners, P.~McFadden, N.~Turok and P.~J. Steinhardt, \emph{{Generating
  ekpyrotic curvature perturbations before the big bang}},
  \href{https://doi.org/10.1103/PhysRevD.76.103501}{\emph{Phys. Rev. D}
  {\bfseries 76} (2007) 103501}
  [\href{https://arxiv.org/abs/hep-th/0702153}{{\ttfamily hep-th/0702153}}].

\bibitem{Ijjas:2014fja}
A.~Ijjas, J.-L. Lehners and P.~J. Steinhardt, \emph{{General mechanism for
  producing scale-invariant perturbations and small non-Gaussianity in
  ekpyrotic models}},
  \href{https://doi.org/10.1103/PhysRevD.89.123520}{\emph{Phys. Rev. D}
  {\bfseries 89} (2014) 123520}
  [\href{https://arxiv.org/abs/1404.1265}{{\ttfamily 1404.1265}}].

\bibitem{Li:2013hga}
M.~Li, \emph{{Note on the production of scale-invariant entropy perturbation in
  the Ekpyrotic universe}},
  \href{https://doi.org/10.1016/j.physletb.2013.06.035}{\emph{Phys. Lett. B}
  {\bfseries 724} (2013) 192}
  [\href{https://arxiv.org/abs/1306.0191}{{\ttfamily 1306.0191}}].

\bibitem{Brandenberger:2020tcr}
R.~Brandenberger and Z.~Wang, \emph{{Nonsingular Ekpyrotic Cosmology with a
  Nearly Scale-Invariant Spectrum of Cosmological Perturbations and
  Gravitational Waves}},
  \href{https://doi.org/10.1103/PhysRevD.101.063522}{\emph{Phys. Rev. D}
  {\bfseries 101} (2020) 063522}
  [\href{https://arxiv.org/abs/2001.00638}{{\ttfamily 2001.00638}}].

\bibitem{Brandenberger:2020eyf}
R.~Brandenberger and Z.~Wang, \emph{{Ekpyrotic cosmology with a zero-shear
  S-brane}}, \href{https://doi.org/10.1103/PhysRevD.102.023516}{\emph{Phys.
  Rev. D} {\bfseries 102} (2020) 023516}
  [\href{https://arxiv.org/abs/2004.06437}{{\ttfamily 2004.06437}}].

\bibitem{Brandenberger:2020wha}
R.~Brandenberger, K.~Dasgupta and Z.~Wang, \emph{{Reheating after S-Brane
  Ekpyrosis}},  \href{https://arxiv.org/abs/2007.01203}{{\ttfamily
  2007.01203}}.

\bibitem{Khoury:2008wj}
J.~Khoury and F.~Piazza, \emph{{Rapidly-Varying Speed of Sound, Scale
  Invariance and Non-Gaussian Signatures}},
  \href{https://doi.org/10.1088/1475-7516/2009/07/026}{\emph{JCAP} {\bfseries
  07} (2009) 026} [\href{https://arxiv.org/abs/0811.3633}{{\ttfamily
  0811.3633}}].

\bibitem{Lehners:2009qu}
J.-L. Lehners and P.~J. Steinhardt, \emph{{Non-Gaussianity Generated by the
  Entropic Mechanism in Bouncing Cosmologies Made Simple}},
  \href{https://doi.org/10.1103/PhysRevD.80.103520}{\emph{Phys. Rev. D}
  {\bfseries 80} (2009) 103520}
  [\href{https://arxiv.org/abs/0909.2558}{{\ttfamily 0909.2558}}].

\bibitem{Fertig:2016czu}
A.~Fertig, J.-L. Lehners, E.~Mallwitz and E.~Wilson-Ewing, \emph{{Converting
  entropy to curvature perturbations after a cosmic bounce}},
  \href{https://doi.org/10.1088/1475-7516/2016/10/005}{\emph{JCAP} {\bfseries
  10} (2016) 005} [\href{https://arxiv.org/abs/1607.05663}{{\ttfamily
  1607.05663}}].

\bibitem{Mukhanov:1990me}
V.~F. Mukhanov, H.~Feldman and R.~H. Brandenberger, \emph{{Theory of
  cosmological perturbations. Part 1. Classical perturbations. Part 2. Quantum
  theory of perturbations. Part 3. Extensions}},
  \href{https://doi.org/10.1016/0370-1573(92)90044-Z}{\emph{Phys. Rept.}
  {\bfseries 215} (1992) 203}.

\bibitem{Brandenberger:2003vk}
R.~H. Brandenberger, \emph{{Lectures on the theory of cosmological
  perturbations}},
  \href{https://doi.org/10.1007/978-3-540-40918-2\_5}{\emph{Lect. Notes Phys.}
  {\bfseries 646} (2004) 127}
  [\href{https://arxiv.org/abs/hep-th/0306071}{{\ttfamily hep-th/0306071}}].

\end{thebibliography}\endgroup



\end{document}